\documentclass[3p,twocolumn,10pt]{scrartcl}
\usepackage[hmarginratio=2:1,top=32mm,left=15mm,columnsep=0.5cm]{geometry}
\usepackage{titling}
\usepackage[svgnames,table]{xcolor} 
\usepackage{tabularx}
\usepackage{graphicx} 
\usepackage[fleqn]{amsmath}
\usepackage{amssymb}
\usepackage{amsfonts}	
\usepackage{moreverb}
\usepackage{dsfont}
\usepackage{tipa}
\usepackage{upgreek}
\usepackage{bm}
\usepackage{multirow}
\usepackage{soul}
\usepackage{ textcomp }
\usepackage{relsize} 
\usepackage{caption}
\usepackage{subcaption}
\usepackage{setspace}
\usepackage{multicol}
\usepackage{wrapfig}

\usepackage{nccmath}

\newcommand\BibTeX{{\rmfamily B\kern-.05em \textsc{i\kern-.025em b}\kern-.08em
T\kern-.1667em\lower.7ex\hbox{E}\kern-.125emX}}

\usepackage[backend=bibtex,giveninits=true,sorting=nyt,url=false,doi=true,eprint=false,isbn=false,
backref,backrefstyle=none,maxbibnames=99]{biblatex}
\DefineBibliographyStrings{english}{%
  backrefpage = {Cited on p\adddot},%
  backrefpages = {Cited on pp\adddot}%
}

\bibliography{library}

\renewbibmacro{in:}{%
  \ifboolexpr{%
     test {\ifentrytype{article}}%
  }{}{\printtext{\bibstring{in}\intitlepunct}}%
}

\usepackage{hyperref} 
\hypersetup{
    colorlinks=true,                          
    linkcolor=blue, 
    citecolor=DarkRed, 
    urlcolor=blue  } 

\newcommand{\pd}{\partial}

\newcommand{\bdm}{\begin{displaymath}}
\newcommand{\edm}{\end{displaymath}}

\newcommand{\bea}{\begin{eqnarray} }
\newcommand{\eea}{\end{eqnarray} }

\renewcommand{\AA}{{\bm{A}}}
\renewcommand{\aa}{{\bm{a}}}
\renewcommand{\ggg}{{\bm{g}}}

\newcommand{\GG}{{\boldsymbol{G}}}

\newcommand{\vv}{{\bm{v}}}

\newcommand{\FF}{{\bm{F}}}

\newcommand{\QQ}{{\bm{Q}}}
\renewcommand{\SS}{{\bm{S}}}

\newcommand{\ww}{{\bm{w}}}

\newcommand{\qq}{{\bm{q}}}

\newcommand{\tr}{\textnormal{tr}}

\newcommand*\samethanks[1][\value{footnote}]{\footnotemark[#1]}
\newcommand{\tort}{{\mathcal{T}}}
\newcommand{\Km}{K_{\textrm m}}
\newcommand{\Ks}{K_{\textrm s}}
\newcommand{\Ku}{K_{\textrm u}}
\newcommand{\Kf}{K_{\textrm f}}
\newcommand{\lambdau}{\lambda_{\textrm u}}
\newcommand{\muu}{\mu_{\textrm u}}
\newcommand{\mus}{\mu_{\textrm s}}
\newcommand{\rhof}{\rho_{\textrm f}}
\newcommand{\rhos}{\rho_{\textrm s}}

\newcommand{\alphaf}{\alpha_{\textrm f}}
\newcommand{\alphas}{\alpha_{\textrm s}}
\newcommand{\cf}{c_{\textrm f}}
\newcommand{\cs}{c_{\textrm s}}
\newcommand{\vf}{v_{\textrm f}}
\newcommand{\vs}{v_{\textrm s}}
\newcommand{\Cs}{C_{\textrm s}}
\newcommand{\Csh}{C_{\textrm sh}}
\newcommand{\Cf}{C_{\textrm f}}
\newcommand{\csh}{c_{\textrm sh}}
\newcommand{\cshm}{c_{{\textrm sh,m}}}

\newfont{\numerikEleven}{ecrm1000}
\newfont{\numerikTen}{cmss10}
\newfont{\numerikNine}{cmss9}
\newfont{\numerikEight}{cmss8}

\title{\textbf{
Modeling wavefields in saturated elastic porous media\\[1mm] based on 
thermodynamically 
compatible 
system theory for two-phase solid-fluid mixtures}}
\author{\fontsize{10pt}{10pt}
	\textsc{Evgeniy Romenski},\hspace{-2mm}
	\thanks{Sobolev Institute of Mathematics, 4 Acad. Koptyug Avenue, 630090 
	Novosibirsk, Russia, 
	\href{mailto:evrom@math.nsc.ru}{e-mail: evrom@math.nsc.ru}
		   }
\quad 
	\textsc{Galina Reshetova},\hspace{-2mm}
	\samethanks[1]
	\hspace{-3mm}
    $\ ^, $
    \hspace{-2.5mm}
 	\thanks{Institute of Computational Mathematics and Mathematical Geophysics, 6 Pr. 
	Akademika Lavrentjeva, 630090 Novosibirsk, Russia, 	\href{mailto:kgv@nmsf.sscc.ru}{e-mail: kgv@nmsf.sscc.ru}
		   }
\quad
	\textsc{Ilya Peshkov},\hspace{-2mm}
	\thanks{Department of Civil, Environmental and Mechanical Engineering, 
	University of Trento, Via Mesiano 77, 38123 Trento, Italy,
	\href{mailto:ilya.peshkov@unitn.it}{e-mail: ilya.peshkov@unitn.it}
		   }$\ ^, $
	   \hspace{-2.5mm}
	   	\thanks{The work by I.P. has been started while being at Paul Sabatier 
	   	University, Institut de 
	   	Mathe\'ematiques 
	   	de Toulouse, Toulouse, France}
\quad
	\textsc{Michael Dumbser}\hspace{-2mm}
	\samethanks[3]
	   }

\thanksmarkseries{arabic}
\allowdisplaybreaks

\begin{document} 
\maketitle

\noindent
\paragraph*{Abstract} \textit{A two-phase model and its application to wavefields numerical 
simulation are 
discussed in the 
context of modeling of compressible fluid flows in elastic porous media. The derivation of the 
model is based on a theory of thermodynamically compatible systems and on a model of nonlinear 
elastoplasticity combined with a two-phase compressible fluid flow model. The governing equations 
of the model include phase mass conservation laws, a total momentum conservation law, an equation 
for the relative velocities of the phases, an equation for mixture distortion, and a balance 
equation for porosity. They form a hyperbolic system of conservation equations that satisfy the 
fundamental laws of thermodynamics. Two types of phase interaction are introduced in the model: 
phase pressure relaxation to a common value and interfacial friction. Inelastic deformations also 
can be accounted for by source terms in the equation for distortion. The thus formulated model can 
be used for studying general compressible fluid flows in a deformable elastoplastic porous medium, 
and for modeling wave propagation in a saturated porous medium. Governing equations for 
small-amplitude wave propagation in a uniform porous medium saturated with a single fluid are 
derived. They form a first-order hyperbolic PDE system written in terms of stress and velocities 
and, like in Biot’s model, predict three type of waves existing in real fluid-saturated porous 
media: fast and slow longitudinal waves and shear waves. For the numerical solution of these 
equations, an efficient numerical method based on a staggered-grid finite difference scheme is 
used. The results of solving some numerical test problems are presented and discussed}


%


%
\section{Introduction} \label{sec:introduction}
The modeling of fluid flows in porous media is of permanent interest in many geophysical and 
industrial applications. The starting point of research developments in this field was a series 
of pioneering works by Biot \cite{Biot1956,Biot1956a,Biot1962}, 
in which a model of elastic wave propagation in 
saturated porous media was proposed. Some modifications and generalizations of 
the model 
have been made (see, for example 
\cite{Carcione2010,Masson2006,Winkler1989} and references therein), and at 
present Biot's approach is 
a commonly accepted and widely used one in geophysical community. Nevertheless, many actual 
technological and scientific problems, such as geothermal energy extraction, CO2 storage, hydraulic 
fracturing, fuel cells, food production, etc. require new advanced models and 
methods. 

To simulate the development of nonlinear, temperature dependent processes in porous media, methods of continuum mechanics and, in particular, multiphase theories can be successfully used. 
A two-phase approach to poroelasticity has been, perhaps, most consistently
implemented by Wilmanski in \cite{Wilmanski1998,Wilmanski2006} (see also 
references therein). In particular, in a review paper, \cite{Wilmanski2006}, 
the structure of Biot's poroelastic model was analysed and its consistency 
with the fundamental principles of continuum mechanics was discussed. 

In recent years, considerable attention has been paid to the modeling 
finite-strain saturated porous media and their applications in various 
fields, 
in particular in medicine, see,
for example, \cite{Khoei2011,Rohan2017,Pesavento2017}.
Worthy of mention is a large-deformation model of a saturated porous 
medium proposed by Dorovsky 
\cite{BlokhinDorovsky1995}, in which the key point of the model is  
thermodynamic consistency and hyperbolicity of the governing equations. 
Nevertheless, there is still no  thermodynamically consistent 
formulation of a multiphase mixture flow model in a  deforming porous medium 
with finite deformations. 

In this paper, we apply a powerful method of designing  new models of 
complex continuum media, which is  based on a thermodynamically compatible system 
theory \cite{God1961,Godunov:1995a,Godunov1996}. 
In \cite{Romenski2013}, the mentioned theory was first time applied to designing a 
model 
for deformed porous medium saturated by a compressible fluid. In this paper the 
governing equations were formulated and small amplitude wave propagation has been studied showing a 
qualitative agreement with Biot's theory. The present study contains further developments of 
ideas from \cite{Romenski2013} based on the unified model of continuum \cite{DPRZ2016,DPRZ2017} as 
well as a quantitative comparison with Biot's theory and a series of numerical experiments proving 
the physical correctness of the model.
The thermodynamically compatible system theory allows the development of 
well-posed models satisfying the
fundamental laws of irreversible thermodynamics. 
In \cite{Godunov:1995a,Rom1998,Romenski2001,Godrom2003,SHTC-GENERIC-CMAT},
a class of Symmetric Hyperbolic Thermodynamically Compatible (SHTC) systems was 
formulated  to describe many known classical equations of 
continuum mechanics and electrodynamics (fluid mechanics, solid mechanics, 
electrodynamics, magnetohydrodynamics)
including advective and dissipative processes. All SHTC systems have nice 
mathematical properties: symmetric hyperbolicity in the sense of Friedrichs 
\cite{Friedrichs1958} and
a conservative form of the equations. The solutions to the governing PDE system 
satisfy 
fundamental laws of non-equilibrium irreversible thermodynamics:  
conservation of total energy (first law) and non-decreasing of  physical 
entropy (second law). 

The SHTC system theory is a first principle type theory. It allows the derivation of
governing PDEs for a quite wide class of physical 
processes from a variational principle \cite{SHTC-GENERIC-CMAT} (Hamilton's 
principle of stationary action). In particular, the SHTC approach has been 
successfully applied to the development of a hierarchy of compressible 
multi-phase flow models 
\cite{Romenski2007,RomDrikToro2010,Romenski2016,PeshGrmRom2015}.
 Recently, a unified SHTC model of Newtonian continuum mechanics has been developed 
\cite{DPRZ2016,DPRZ2017}. It simultaneously describes the dynamics of 
elastoplastic solids, as well as of viscous and non-viscous fluids in the presence 
of electromagnetic fields. 
In the present paper, we extend this unified model to  describe 
solid-fluid two-phase flows. The governing equations of the model 
also belong to the class of SHTC equations. The interfacial friction between the
liquid and solid phases and shear stress relaxation are implemented in the 
model as relaxation-type source terms in accordance with the laws of 
thermodynamics. 
The latter allows taking into account the time dependence of elastic moduli on frequency and using numerical methods to study wave 
propagation problems.  

The rest of the paper is organized as follows. Section\,\ref{sec.theory} 
briefly describes the governing PDEs for a unified model of continuum mechanics 
and for a
two-phase compressible fluid model. By combining the above two  
models, a governing master SHTC system for a two-phase solid-fluid medium is 
formulated.
In Section\,\ref{sec.linear}, using this model governing 
equations are derived for small-amplitude wave propagation in a stationary 
saturated porous medium. 
In Section \ref{Comparison} we
discuss the differences and similarities between the Biot and SHTC models.
In Section\,\ref{sec.waves}, we derive a dispersion 
relation for the thus obtained acoustic equations and study the properties of the wavefields. 
In Section\,\ref{sec.numerics}, an efficient finite difference method for 
small-amplitude wave propagation is presented and some numerical results are 
discussed.

%

\section{System of governing PDEs for poroelastic media}\label{sec.theory}

To develop a poroelastic model whose governing equations form a symmetric 
hyperbolic thermodynamically compatible (SHTC) system, a unified thermodynamically compatible continuum model formulated in 
\cite{HPR2016,DPRZ2016} is coupled to a
two-phase compressible fluid model \cite{RomDrikToro2010}. 

\subsection{SHTC governing equations of the unified model of 
continuum}\label{sec.GPR}

We start our consideration with a brief recall of the unified model of continuum, the 
formulation of which can be found in \cite{HPR2016,DPRZ2016}. This model can describe elastic and
elastoplastic behavior of the medium, as well as flows of a viscous fluid by a single PDE system.
First, we recall the governing equations of the unified model for a deforming 
and flowing continuum which reads
\begin{subequations}\label{eqn.HPR}
\begin{eqnarray} 
	&&\displaystyle\frac{\partial \rho v^i}{\partial t}+\frac{\partial 
		\left(\rho v^i v^k + p \delta_{ik} - \sigma_{ik} \right)}{\partial x_k}=0, 
	\label{eqn.momentum}\\[2mm]
	&& \frac{\partial \rho}{\partial t}+\frac{\partial \rho v^k}{\partial 
	x_k}=0,\label{eqn.conti}\\[2mm]
	&&\displaystyle\frac{\partial A_{i k}}{\partial t}+\frac{\partial A_{ij} 
		v^j}{\partial x_k}+v^j\left(\frac{\partial A_{ik}}{\partial 
		x_j}-\frac{\partial A_{ij}}{\partial x_k}\right)
	=-\dfrac{ \psi_{ik} }{\theta},\label{eqn.deformation}\\[2mm]
	&&\displaystyle\frac{\partial \rho s}{\partial t}+\frac{\partial \rho 
		s v^k }{\partial x_k}=\dfrac{\rho}{\theta T} 
	\psi_{ik} \psi_{ik} \geq0. 
	\label{eqn.entropy}
\end{eqnarray}
\end{subequations}
Here, \eqref{eqn.momentum} is the linear momentum conservation law, 
\eqref{eqn.conti} 
is the mass conservation law, \eqref{eqn.deformation} is the evolution for the
distortion matrix, and \eqref{eqn.entropy} is the entropy balance law.

As an independent set of state parameters we take the medium
velocity $v^k$, mass density $\rho$, distortion $A_{ik}$, and entropy $s$. Note that 
positioning of the tensorial index in the velocity field $ v^k $ as a superscript is merely 
conditioned by 
that the subscript is reserved for the constituent index and is not intending to distinguish 
covariant and contravariant components of a tensor which coincide with each other in a 
Cartesian coordinate system.
The density is connected with the distortion  by an algebraic compatibility constraint such as 
$\rho=\rho_0 \det\hspace{-0.6mm}\AA$, where $\rho_0$ is a reference density.
Additional parameters of the medium presented in the above system are 
pressure $p$, shear stress $\sigma_{ik}$, and temperature $T$. They are 
connected with density, distortion, and entropy via specific total 
energy 
$E(\vv,\rho,s,\AA)$:
\begin{align}
& p:=\rho^2\frac{\partial E}{\partial \rho} \equiv \rho^2 E_\rho, 
\  
\sigma_{ij} :=-\rho A_{ki}\frac{\partial E}{\partial A_{kj}} \equiv -\rho 
A_{ki} E_{ 
A_{kj}}, 
\notag\\ 
& T:=\frac{\partial E}{\partial s} \equiv E_s.
\end{align}
The source term in the equation for distortion characterizes the rate of inelastic deformation,
where $\bm{\psi}=\frac{\partial E}{\partial \AA}$, while the function $\theta(\tau)$ depends on the 
shear stress relaxation time $\tau$. This strain relaxation source term is one of the 
key elements of the model since it makes it possible to apply the model to description of both 
elasto-plastic deformations of solids \cite{Godrom2003,Hyper-Hypo2019} as well as flows of viscous 
fluids \cite{DPRZ2016,HYP2016,Jackson2019a,Busto2019}.

To close the model, we define how the energy $E$ and the parameter $\theta$ depend on the parameters of state.
We take the energy in the form
\begin{align}
E=E_1(\rho,s)+E_2(\rho,s,\AA)+E_3(\vv), \label{energy}
\end{align}
where $E_3=\frac{1}{2}\Vert \vv \Vert^2$ is specific kinetic energy,
$E_1$ is the "hydrodynamic" part corresponding to the energy of  volume deformations only,
and $E_2$ is the energy of shear strain.
In order to provide a zero trace of shear stress, $\tr (\bm{\sigma})=0$, we take 
$E_2$ depending on $\AA$ via the normalized strain tensor $\ggg=\aa^T\aa$, 
where 
$\aa=\AA/(\det\hspace{-0.6mm}\AA)^{1/3}$. It gives  
$\ggg=\GG/(\det\hspace{-0.4mm}\GG)^{1/3}$, where 
$\GG=\AA^T\AA$. Then, energy $ E_2 $ can be 
defined as in \cite{Ndanou2014}:
\begin{align}
E_2=\frac{1}{8}\csh^2\left(\tr({\ggg^2})-3\right),
\end{align}
where $\csh$ is the shear velocity of sound under reference conditions.

Using the above definition, we can take
the derivative of $E$ with respect to $\AA$:
\begin{align} \label{EderivA}
\frac{\partial E}{\partial \AA}=\frac{\partial E_2}{\partial 
\AA}=-\rho^{-1}\FF^T\bm{\sigma}=
\frac{\csh^2}{2} \bm{F}^T \left(\ggg^2-\frac{\tr({\ggg^2})}{3} \bm{I} \right),
\end{align}
where $\FF=\AA^{-1}$.
Then the shear stress is trace-free, and it reads as
\begin{align}
\bm{\sigma} =-\rho \frac{\csh^2}{2}\left(\ggg^2-\frac{\tr({\ggg^2})}{3} \bm{I}
\right), \quad  \tr(\bm{\sigma})=0.
\end{align} 

The coefficient $\theta$ is a function of parameters of state, and it can 
be taken in the form \cite{DPRZ2016,HYP2016,Hyper-Hypo2019}
\begin{align}
\theta \sim \csh^2 \tau (\rho,T,Y) \textgreater 0,
\end{align}
where $Y=\sqrt{\frac32\tr(\bm{\sigma}^2)}$ is the intensity of shear stress.

The source term in the equation for distortion produces a nonnegative entropy production source term in \eqref{eqn.entropy}. 
It is important to note that for system \eqref{eqn.HPR} there is an additional 
conservation law \cite{SHTC-GENERIC-CMAT}:
\begin{align} \label{eqn.energy}
\displaystyle\frac{\partial \rho E}{\partial t}+
\frac{\partial \left(\rho  v^k E +v^i(p \delta_{ik}-\sigma_{ik}) 
\right)}{\partial x_k}=0,
\end{align}
which is the conventional energy conservation law.

\subsection{SHTC governing equations of two-phase compressible fluid flow model}
\label{sec.twophase}

In the two-phase compressible fluid model, the flow is considered  
as a mixture of two 
immiscible constituents with their own parameters of state. Thus, the general model should take into 
account the difference between velocities, pressures, and temperatures of the 
constituents. That means that if 
we consider  velocity, density, and entropy to be the basic parameters of 
state, they can be 
different for each phase. 
In our consideration, we restrict ourselves to a single entropy approximation for small variations 
of phase temperatures. In \cite{Romenski2016}, it is proved that 
for multiphase compressible mixtures a single entropy approximation is suitable for the flows 
close to thermal equilibrium. In this case, the change of phase temperatures due to the small 
variation of phase entropies is negligibly small.
Finally note, that we consider only small density variations which means that the processes under 
consideration are close to thermal equilibrium.

The SHTC system for two-phase compressible flow with a single entropy approximation 
\cite{RomDrikToro2010} reads as follows:
\begin{subequations}\label{eqn.HPRFF}
\begin{fleqn}[0pt]
\begin{eqnarray}
&&\displaystyle\frac{\partial \rho v^i}{\partial t}+\frac{\partial 
	(\rho v^i v^k + p \delta_{ik} + w^iE_{w^k} )}{\partial x_k}=0, 
\label{eqn.momentumFF}\\[2mm]
&& \frac{\partial \rho}{\partial t}+\frac{\partial \rho v^k}{\partial 
	x_k}=0,\label{eqn.contiFF}\\[2mm]
&& \frac{\partial \rho c_1}{\partial t}+\frac{\partial (\rho c_1 v^k+\rho E_{w_k})}{\partial 
	x_k}=0,\label{eqn.contiFF1}\\[2mm]
&&\displaystyle\frac{\partial w^k}{\partial t}+\frac{\partial(w^lv^l+E_{c_1})}{\partial x_k}+v^l\left(\frac{\partial w^k}{\partial x_l}-
\frac{\partial w^l}{\partial x_k}\right)=-\dfrac{ \lambda_{k} }{\theta_2},\label{eqn.relvel}\\[2mm]
&& \frac{\partial \rho \alpha_1}{\partial t}+\frac{\partial \rho \alpha_1 v^k }{\partial x_k}=-\frac{\rho \phi}{\theta_1},\label{eqn.alphaFF}\\[2mm]
&&\displaystyle\frac{\partial \rho s}{\partial t}+\frac{\partial \rho 
	s v^k }{\partial x_k}=\dfrac{\rho}{\theta_1 T}\phi^2 +
\dfrac{\rho}{\theta_2 T}\lambda_k \lambda_k \geq0, 
\label{eqn.entropyFF}
\end{eqnarray}
\end{fleqn}
\end{subequations}
Here, $\alpha_1$ is the volume fraction of the first phase, which is connected 
with the volume 
fraction of the second phase $\alpha_2$ via the saturation law 
$\alpha_1+\alpha_2=1$, $\rho$ is the mixture mass density, which is connected 
with the phase mass densities $\rho_1,\rho_2$ via the relation 
$\rho=\alpha_1\rho_1+\alpha_2\rho_2$. The phase mass fractions are defined as 
$c_1=\alpha_1 \rho_1/\rho, c_2=\alpha_2 \rho_2/\rho$  $(c_1+c_2=1)$. 
Eventually, 
$v^i=c_1v_1^i+c_2v_2^i$ is the mixture velocity, $w^i=v_1^i-v_2^i$ is the 
phase relative velocity, $s$ is the specific entropy of the mixture, and as 
throughout this paper, we use the notations $ E_{w^k} \equiv \frac{\pd E}{\pd 
w^k} $, $ E_{c_1} 
\equiv \frac{\pd E}{\pd c_1} $, etc.

The phase interaction is presented as algebraic source terms in 
\eqref{eqn.relvel} and \eqref{eqn.alphaFF} which are proportional to 
thermodynamic forces. These source terms are phase pressure relaxation to a 
common value and interfacial friction: $-\rho{\phi}/{\theta_1} = 
-\rho{E_{\alpha_1}}/{\theta_1}$ and $-{\lambda_{k} }/{\theta_2} = 
-{E_{w^k}}/{\theta_2}$. The coefficients $\theta_1,\theta_2$ characterize the rate 
of pressure and the velocity relaxation. They can depend on the parameters of state.

The entropy production in the entropy balance equation \eqref{eqn.entropyFF} is 
non-negative due to the definition of the phase interaction source terms.
The energy conservation law holds in the following form:
\begin{align} \label{eqn.energyFF}
\displaystyle\frac{\partial \rho E}{\partial t}+
\frac{\partial \left(\rho  v^k E +v^i(p \delta_{ik}+\rho w^i E_{w^k}) +\rho E_{c_1}E_{w_k} \right)}{\partial x_k}=0,
\end{align}
where $E=E_1(\alpha_1, c_1, \rho, s)+E_3(\vv)+E_4(c_1,\ww)$, $E_1$ is the 
specific internal energy of the mixture, $E_3=\Vert\vv\Vert^2/2$ is the kinetic energy of 
the mixture, and $E_4=\frac{1}{2}c_1c_2\Vert\ww\Vert^2$ is the kinematic energy of the
relative motion.
Note that 
\begin{align}
E_3+E_4=\frac{1}{2}(\alpha_1\rho_1 \Vert\vv_1\Vert^2+\alpha_2\rho_2 \Vert\vv_2\Vert^2). 
\end{align}

The most important closing relation for system \eqref{eqn.HPRFF}
is the internal energy  $E_1$, which should be chosen in such a way that the 
governing equations  take the form of the well-known balance laws of the two-phase  
flow model. In \cite{RomDrikToro2010,Romenski2016} it is defined as the mass 
averaged phase equation of state
\begin{align}
E_1=c_1e_1(\rho_1,s)+c_2e_2(\rho_2,s),
\end{align} 
where $e_i(\rho_i,s)$ is the specific internal energy of the $ i $-th phase.

This definition leads to the following formulae for the thermodynamic forces (the derivatives of internal energy with respect to the parameters of state) \cite{RomDrikToro2010,Romenski2016}:
\begin{align} \label{Thermod.forces}
& E_{\alpha_1}=\frac{p_2-p_1}{\rho},\ \  p=\rho^2E_\rho={\alpha_1p_1+\alpha_2p_2},\ \  
E_{w_i}=c_1c_2 w_i, \notag\\
& E_{c_1}=e_1+p_1/\rho_1-e_2-p_2/\rho_2+(1-2c_1)\Vert\ww\Vert^2/2,\notag\\ 
& E_s=T=c_1\frac{\partial e_1}{\partial s}+c_2\frac{\partial e_2}{\partial s}. 
\end{align}  
Here $p_i=\rho_i^2\frac{\partial e_i}{\partial \rho_i}, (i=1,2)$ is the phase 
pressure.

In \cite{Romenski2016}, a model of multiphase compressible flow with 
arbitrary number of phases is presented, and it is shown that its governing 
equations can be written as a symmetric hyperbolic system.
It means that the two-phase equations \eqref{eqn.HPRFF} with the closing 
relations \eqref{Thermod.forces} can also be written as a symmetric hyperbolic 
system.

\subsection{Thermodynamically compatible master system for two-phase saturated porous media}

The SHTC systems of the unified continuum model and the two-phase compressible fluid flow model presented in 
the previous sections can be obtained as subsystems of a general master 
system written with the use 
of a generalized internal energy. This master system can 
be derived from the first principles by 
minimizing the Lagrangian and passing to  Eulerian coordinates, see 
\cite{SHTC-GENERIC-CMAT}.  Its derivation and connection with a Hamiltonian 
formulation of irreversible non-equilibrium thermodynamics known as GENERIC is 
discussed in \cite{SHTC-GENERIC-CMAT}. In 
\cite{PeshGrmRom2015}, the above-mentioned master system is applied to the derivation 
of a model for 
two-phase solid-fluid media experiencing a stress-induced solid-fluid phase 
transformation. The goal of the present 
paper is to develop a two-phase model for compressible fluid flow in deforming porous media.

Let us consider the following PDE master system:
\begin{subequations}\label{eqn.MS}
	\begin{eqnarray}
&&\displaystyle\frac{\partial \rho v^i}{\partial t}+\frac{\partial 
	(\rho v^i v^k + p \delta_{ik} + w^iE_{w^k} -\sigma_{ik} )}{\partial x_k}=0, 
\label{eqn.momentumMS}\\[2mm]
	&&\displaystyle\frac{\partial A_{i k}}{\partial t}+\frac{\partial A_{im} 
	v^m}{\partial x_k}+v^j\left(\frac{\partial A_{ik}}{\partial 
	x_j}-\frac{\partial A_{ij}}{\partial x_k}\right)
=-\dfrac{ \psi_{ik} }{\theta},\label{eqn.deformationMS}\\[2mm]
&& \frac{\partial \rho}{\partial t}+\frac{\partial \rho v^k}{\partial 
	x_k}=0,\label{eqn.contiMS}\\[2mm]
&& \frac{\partial \rho c_1}{\partial t}+\frac{(\partial \rho c_1 v^k+\rho 
E_{w^k})}{\partial 
	x_k}=0,\label{eqn.contiMS1}\\[2mm]
&&\displaystyle\frac{\partial w^k}{\partial t}+\frac{\partial (w^lv^l+E_{c_1})}{\partial 
x_k}
+v^l\left(\frac{\partial w^k}{\partial x_l}-
\frac{\partial w^l}{\partial x_k}\right)
=-\dfrac{ \lambda_{k} }{\theta_2},\label{eqn.relvelMS}\\[2mm]
&& \frac{\partial \rho \alpha_1}{\partial t}+\frac{\partial \rho \alpha_1 v^k 
}{\partial 
	x_k}=-\frac{\rho \varphi}{\theta_1},\label{eqn.alphaMS}\\[2mm]
&&\displaystyle\frac{\partial \rho s}{\partial t}+\frac{\partial \rho 
	s v^k }{\partial x_k}=\dfrac{\rho}{\theta T}\psi_{ik} \psi_{ik}+
\dfrac{\rho}{\theta_1 T}\varphi^2 +
\dfrac{\rho}{\theta_2 T}\lambda_k \lambda_k \geq0, 
\label{eqn.entropyMS}
\end{eqnarray}
\end{subequations}
   
The application of the master system \eqref{eqn.MS} to designing a concrete 
physical process has to 
be done by the proper choice of the generalized energy potential $ E $. If, for 
example, we take $E=E(\vv,\rho,\AA,s)$, we can neglect equations 
\eqref{eqn.contiMS1}, \eqref{eqn.relvelMS}, \eqref{eqn.alphaMS} and obtain
the PDEs for the unified model of continuum mechanics \eqref{eqn.HPR}.
On the other hand, if we take $E=E(\alpha_1, c_1, \rho, \ww, s)$, we 
obtain the 
governing PDEs for compressible two-phase flow \eqref{eqn.HPRFF}. Therefore,
the governing PDEs for deformed continuum and for compressible two-phase fluid 
flow can be viewed as consequences of \eqref{eqn.MS}. It seems natural to 
take this general system as a basis for a two-phase solid-fluid mixture model.

First, we identify the parameters of state in \eqref{eqn.MS} with 
physical parameters characterizing 
deforming porous media.
Let the parameter $\alpha_1$ characterize the volume fraction of the fluid component 
in the solid-fluid mixture, which means that it can be identified with the 
porosity,
usually denoted by $ \phi $. Then $\alpha_2=1-\alpha_1$ is the volume fraction 
of the solid phase of the porous material. As in Section\,\ref{sec.twophase}, the mixture density  
$\rho$ is connected 
with the phase mass densities via $\rho=\alpha_1\rho_1+\alpha_2\rho_2$. 
The parameter 
$c_1$ represents the mass fraction of the fluid component. And, by analogy 
with the two-phase flow
model, 
\begin{equation}\label{eqn.vel}
	v^i=c_1v_1^i+c_2v_2^i, \qquad  w^i=v_1^i-v_2^i 
\end{equation}
is the velocity of the mixture and the relative velocity, respectively.
We consider a single entropy approximation of the two-phase medium and introduce 
the entropy of the mixture $s$. 
As noted at the beginning of Section 2.2, for processes which are close to thermal 
equilibrium, the single entropy approximation is acceptable.
Finally, as a measure of deformation of the 
element of the porous medium we consider the distortion $\AA$ of the mixture.

Let us now take the generalized total energy potential in the same form as in 
the case 
of a unified deforming continuum \eqref{energy}, but taking into account 
the two-phase nature of the medium and assuming that the energy corresponding to  
volume deformation is defined as in the two-phase fluid model: 
\begin{align}
E=E_1(\alpha_1, c_1, \rho, s)+E_2(c_1,\rho,s,\AA)+E_3(\vv)+E_4(c_1,\ww). 
\label{energy.SF}
\end{align}
Here we assume that 
\begin{align} \label{energy12.SF}
&E_1=c_1e_1(\rho_1,s)+c_2e_2(\rho_2,s), \quad
E_3=\frac{1}{2}\Vert\vv\Vert^2, \notag\\
& E_4=c_1c_2\frac{1}{2}\Vert\ww\Vert^2.
\end{align}
We take the part of energy related to volume deformation as the mass 
averaged energy of the phases, because it naturally follows from the additivity 
of the energy per unit volume:
\begin{align} \label{energy.mix}
\rho E_1=\alpha_1 \rho_1e_1(\rho_1,s)+\alpha_2 \rho_2 e_2(\rho_2,s). 
\end{align}
The kinetic energy of the relative motion, $E_4$, remains the same as is in the 
two-phase flow model.

There is no rigorous justification for determining the shear strain energy, $ E_2 $. It is clear 
that the volume deformation of two-phase mixtures can be represented as the sum of the volume 
deformations of the solid and fluid constituents, that is why we define the volumetric energy as 
the mass averaged phase energies. As for any arbitrary strain measure of the mixture element, it is 
not clear whether it can be divided into parts related to the
solid and fluid constituents separately. That is why we use the distortion 
of the 
mixture as a parameter of state, the rate of which is connected with spatial derivatives of the 
mixture velocity. Moreover,
we assume that the shear part of internal energy depends on the elastic deformation tensor of the 
mixture and define it by formula   \eqref{sec.GPR}:
$
E_2=\frac18 \cshm^2\left(\tr({\ggg^2})-3\right), 
$
where $\cshm$ is the shear speed of sound of the mixture to be determined.
It can be naturally assumed that in the limiting case $ c_1 = 0, c_2 = 1 $, which corresponds to 
the pure solid state, the shear energy corresponds to the elastic shear energy if $\cshm$ equals 
the 
shear sound 
velocity of the solid medium. On the other hand, recalling that we consider the ideal fluid (no 
resistance 
to shear), we can take $\cshm=0 $ in the pure fluid limiting case $(c_1=1, c_2=0)$ and hence, there 
is no 
contribution to the shear part of the total energy. From this reasoning, it follows that for the 
intermediate case $(0 < c_1,c_2 < 1,)$, we should choose $\cshm$ in such a way that it disappears 
for 
the pure fluid medium and is equal to the solid shear speed of sound for the pure solid medium. 
That is 
why we assume that for the solid-fluid mixture $\cshm^2=c_2\csh^2$, where $\csh$ is the shear speed 
of 
sound of the skeleton, and take the shear energy in the form
\begin{align}\label{energy,shear}
E_2=c_2 \frac18\csh^2\left(\tr({\ggg^2})-3\right). 
\end{align}
It can be shown then via the characteristic analysis \cite{Romenski2013} that such a choice of the 
shear elastic energy gives the shear characteristic speed for the mixture $\cshm=\sqrt{c_2}\csh$, 
that seems to be a reasonable approximation but can be further refined if it is necessary to 
achieve 
better agreement with experimental data.

As soon as we define the generalized energy by \eqref{energy.SF}, 
\eqref{energy12.SF}, \eqref{energy.mix}, \eqref{energy,shear}, the closing 
relations in the phase interaction source terms are defined via the 
thermodynamic 
forces:
\begin{align}
\psi_{ik}=E_{A_{ik}}, \quad \lambda_{k}=E_{w_k}, \quad \varphi=E_{\alpha_1}.
\end{align} 

The solution to system \eqref{eqn.MS} satisfies the energy conservation law 
which reads as
\begin{multline} \label{eqn.energyMS}
\displaystyle\frac{\partial \rho E}{\partial t}+
\frac{\partial}{\partial x_k}\left(\rho  v^k E +v^i(p \delta_{ik} +\rho w^i E_{w^k}-\sigma_{ik}) 
\right .\\
\left. 
+\rho E_{c_1}E_{w_k} \right)=0
\end{multline}
and should be used instead of the entropy balance law \eqref{eqn.entropyMS} in the development of numerical methods. 

With the above definitions of the energies, we obtain the 
following system of governing PDEs for compressible flow in deforming 
porous media, which is written in terms of the phase parameters:
\begin{subequations}\label{eqn.PV}
	\begin{eqnarray}
&&\displaystyle\frac{\partial 
(\alpha_1\rho_1v^i_1+\alpha_2\rho_2v^i_2)}{\partial t}+
\frac{\partial}{\partial x_k} \left (\alpha_1\rho_1v^i_1v^k_1+\alpha_2\rho_2v^i_2v^k_2 \right.
\notag\\
&&\left. \qquad \qquad \qquad \qquad \qquad \qquad \quad \phantom{v^k_2} +p\delta_{ik}-\sigma_{ik}
\right)=0, 
\label{eqn.momentumPV}\\[2mm]
&&\displaystyle\frac{\partial A_{i k}}{\partial t}+\frac{\partial A_{ij} 
	v^j}{\partial x_k}+v^j\left(\frac{\partial A_{ik}}{\partial 
	x_j}-\frac{\partial A_{ij}}{\partial x_k}\right)
=-\dfrac{ \psi_{ik} }{\theta},\label{eqn.deformationPV}\\[2mm]
&& \frac{\partial \alpha_1\rho_1}{\partial t}+\frac{\partial \alpha_1\rho_1 
v^k_1}{\partial 
	x_k}=0,\label{eqn.contiPV1}\\[2mm]
&& \frac{\partial \alpha_2\rho_2}{\partial t}+\frac{\partial \alpha_2\rho_2 
v^k_2}{\partial 
	x_k}=0,\label{eqn.contiPV2}\\[2mm]
&&\displaystyle\frac{\partial w^k}{\partial 
t}+\frac{\left (\frac12(v_1^j v_1^j-v_2^j 
v_2^j)+e_1+\frac{p_1}{\rho_1}-e_2-\frac{p_2}{\rho_2}-E_2\right )}{\partial 
x_k} \notag\\
&&\qquad \qquad \qquad \qquad \qquad+v^l\left(\frac{\partial w^k}{\partial x_l}-
\frac{\partial w^l}{\partial x_k}\right)
=-\dfrac{ \lambda_{k}
}{\theta_2}, \label{eqn.relvelPV}\\[2mm]
&& \frac{\partial \rho \alpha_1}{\partial t}+\frac{\partial \rho \alpha_1 v^k 
}{\partial 
	x_k}=-\frac{\rho \varphi}{\theta_1},\label{eqn.alphaPV}\\[2mm]
&&\displaystyle\frac{\partial \rho s}{\partial t}+\frac{\partial \rho 
	s v^k }{\partial x_k}=\dfrac{\rho}{\theta T}\psi_{ik} \psi_{ik}+
\dfrac{\rho}{\theta_1 T}\varphi^2 +
\dfrac{\rho}{\theta_2 T}\lambda_k \lambda_k \geq0.
\label{eqn.entropyPV}
\end{eqnarray}
\end{subequations}
For the derivation of the above system from equations \eqref{eqn.MS}, the following 
formulae for the thermodynamic forces are used:
\begin{align} \label{Thermod.forcesSF}
& E_{\alpha_1}=\frac{p_2-p_1}{\rho}, \ \  p=\rho^2E_\rho={\alpha_1p_1+\alpha_2p_2}, \notag \\  
&\sigma_{ij}=\alpha_2 s_{ij}, \ \ s_{ij}=- 
\frac{\rho_2\csh^2}{2}({g_{ik}g_{kj}-\frac{1}{3}{g_{mn}g_{nm}}\delta_{ij}}),
\quad  \nonumber \\
&E_{w^i}=c_1c_2 w^i, \notag \\ 
& 
E_{c_1}=e_1+\frac{p_1}{\rho_1}-e_2-\frac{p_2}{\rho_2}-c_2^{-1}E_2+(1-2c_1)\frac{\Vert\ww\Vert^2}{2},
 \notag\\
& E_2=c_2 \frac{1}{8}\csh^2\left(\tr({\ggg^2})-3\right), \notag\\
&  
E_s=T=c_1\frac{\partial e_1}{\partial s}+c_2\frac{\partial e_2}{\partial s}.
\end{align}  
For the derivation of \eqref{Thermod.forcesSF} we use formula \eqref{energy.mix} for the internal energy of the mixture and the relationship between the mixture parameters and the individual phase densities: $\rho_1=\frac{\rho c_1}{\alpha_1}, \rho_2=\frac{\rho c_2}{\alpha_2}$.
Details of the derivation can be found, for example, in 
\cite{RomDrikToro2010,Romenski2016}. 
 
System \eqref{eqn.PV} is equivalent to \eqref{eqn.MS} if the
total mass conservation law 
\eqref{eqn.contiMS} is replaced by the mass conservation law for the second phase 
\eqref{eqn.contiPV2}.
 
\section{System of governing PDEs for small-amplitude wave propagation in  
saturated porous media}\label{sec.linear}

In this section, we derive equations for small-amplitude wave 
propagation in a saturated porous medium at equilibrium. For the 
derivation, we first transform equations \eqref{eqn.PV} to a more convenient form. 
Instead of the total momentum equation 
and the equation for the relative velocity, consider momentum equations for each of the
constituents. These are derived from \eqref{eqn.momentumPV} and 
\eqref{eqn.relvelPV} and read as
\begin{eqnarray}\label{eqns.momentum1.quasilinear}
&&\frac{\partial v^i_1}{\partial t}+v^k_1\frac{\partial v^i_1}{\partial x_k}+
\frac{1}{\rho_1}\frac{\partial p_1}{\partial x_i}- \frac{1}{\rho}
\frac{\partial \alpha_2 s_{ik}}{\partial x_k}+\frac{p_1-p_2}{\rho}\frac{\partial \alpha_1}{\partial 
x_i} \notag\\
&&-c_2 a_{mn}\frac{\partial A_{mn}}{\partial x_i}
+c_2(v^k_1-v^k_2)\left(c_2\left(\frac{\partial v^i_1}{\partial x_k}-\frac{\partial v^k_1}{\partial 
x_i}\right) \right.\notag\\
&& \left. + c_1\left(\frac{\partial v^i_2}{\partial x_k}-\frac{\partial v^k_2}{\partial x_i}\right) 
\right)
=-\frac{\alpha_2 \rho_2}{\rho} \frac{c_1c_2}{\theta_2}(v_1^i-v_2^i),
\end{eqnarray}
\begin{eqnarray} \label{eqns.momentum2.quasilinear}
&&\frac{\partial v^i_2}{\partial t}+v^k_2\frac{\partial v^i_2}{\partial x_k}+
\frac{1}{\rho_2}\frac{\partial p_2}{\partial x_i}- \frac{1}{\rho}
\frac{\partial \alpha_2 s_{ik}}{\partial x_k}+\frac{p_1-p_2}{\rho}\frac{\partial \alpha_1}{\partial 
x_i} \notag\\
&& +c_1 a_{mn}\frac{\partial A_{mn}}{\partial x_i}
-c_1(v^k_1-v^k_2)\left(c_2\left(\frac{\partial v^i_1}{\partial x_k}-\frac{\partial v^k_1}{\partial 
x_i}\right)\right. \notag\\ 
&& \left. + c_1\left(\frac{\partial v^i_2}{\partial x_k}-\frac{\partial v^k_2}{\partial 
x_i}\right) \right)
=\frac{\alpha_1 \rho_1}{\rho} \frac{c_1c_2}{\theta_2}(v_1^i-v_2^i),
\end{eqnarray}
where $a_{mn}={\partial (c_2^{-1}E_2)}/{\partial A_{mn}}$ and the derivative ${\partial E_2}/{\partial A_{mn}}$ can be taken as in \eqref{EderivA}.

In addition to the above equations, we consider phase mass conservation 
laws 
and balance equations for the volume fraction, distortion, and entropy, which can be 
written in the following quasilinear form:
\begin{subequations}\label{eqns.quasilinear}
	\begin{eqnarray} 
&&\frac{\partial \rho_1}{\partial t}+v^k_1\frac{\partial \rho_1}{\partial 
x_k}+\rho_1 \frac{\partial v^k_1}{\partial x_k}+
\frac{\rho_1c_2}{\alpha_1}(v^k_1-v^k_2)\frac{\partial \alpha_1}{\partial x_k} \notag\\
&& = \frac{\rho_1}{\alpha_1 \rho} \frac{p_2-p_1}{\theta_1},   \\[2mm]
&&\frac{\partial \rho_2}{\partial t}+v^k_2\frac{\partial \rho_2}{\partial 
x_k}+\rho_2 \frac{\partial v^k_2}{\partial x_k}+
\frac{\rho_2c_1}{\alpha_2}(v^k_1-v^k_2)\frac{\partial \alpha_1}{\partial x_k} \notag\\
&& = -\frac{\rho_2}{\alpha_2 \rho} \frac{p_2-p_1}{\theta_1},   \\[2mm]
&&\frac{\partial \alpha_1}{\partial t}+v^k\frac{\partial \alpha_1}{\partial 
x_k}=
\frac{p_1-p_2}{\rho \theta_1},  \\[2mm]
&&\frac{\partial A_{i j}}{\partial t}+v^k \frac{\partial A_{ij}}{\partial x_k}
+A_{im}\frac{\partial v^m}{\partial x_j}
=-\dfrac{ E_{A_{ik}} }{\theta},  \\[2mm]
&&\frac{\partial s}{\partial t}+v^k\frac{\partial s}{\partial x_k}=
\dfrac{1}{\theta T} E_{A_{ik}}E_{A_{ik}} +
\dfrac{1}{\theta_1 T}\frac{(p_1-p_2)^2}{\rho} \notag\\ 
&&+
\dfrac{1}{\theta_2 T} E_{w^k}E_{w^k}.  
\end{eqnarray}

\end{subequations}

Small amplitude wave propagation can be described by a PDE system obtained by 
linearization of \eqref{eqns.quasilinear} with coefficients defined in 
the equilibrium state of the original system. Assume that a medium with 
given volume fractions of the constituents is at rest and under reference 
conditions. This means that its parameters of state are
\begin{align}  \label{static.solution}
&v_1^k=v_2^k=0, \quad \rho_1=\rho_1^0, \quad \rho_2=\rho_2^0, 
\quad \alpha_1=\alpha_1^0, \notag\\ 
&\alpha_2=\alpha_2^0 = 1-\alpha_1^0, \quad A_{ij}=\delta_{ij}, \quad s=0.  
\end{align}  
The other parameters of the medium computed with the above ones also correspond to the medium at rest:
\begin{align}
p_1=p_2=0, \quad s_{ik}=0, \quad v^k=0, \quad w^k=0, \quad a_{mn}=0.
\end{align}

We are interested in differential equations for small 
perturbations of the equilibrium solution \eqref{static.solution}. Thus, our 
goal is to find a solution  to the system of equations \eqref{eqns.momentum1.quasilinear}, 
\eqref{eqns.momentum2.quasilinear}, and 
\eqref{eqns.quasilinear} in the form
\begin{subequations}\label{perturbed.solution} 
	\begin{gather} 
		v_1^k=0 +\delta v_1^k, \quad v_2^k=0 +\delta v_2^k, \quad
		\rho_1=\rho_1^0 +\delta \rho_1, \notag \\
		 \rho_2=\rho_2^0+\delta \rho_2, 
		\\[2mm] 
		\alpha_1=\alpha_1^0+\delta \alpha_1, \quad A_{ij}=\delta_{ij}+\delta 
		A_{ij}, \quad s=0+\delta s.
	\end{gather}
\end{subequations}
Equations for perturbations of the equilibrium solution can be derived by 
substituting \eqref{perturbed.solution} into 
\eqref{eqns.momentum1.quasilinear}, 
\eqref{eqns.momentum2.quasilinear}, and 
\eqref{eqns.quasilinear} and neglecting the terms of orders higher than the 
first one. 
The following relations for perturbed phase pressures will also be used:
\begin{align}\label{pressure.linear}
\delta p_1 =\frac{K_1}{\rho_1^0} \delta \rho_1, \quad \delta p_2 = \frac{K_2}{\rho_2^0} \delta \rho_2,
\end{align}
where $
K_1=\left. \rho_1^0\frac{\partial p_1}{\partial \rho_1}\right|_{\rho_1=\rho_1^0, s=0},
\quad
K_2=\left. \rho_2^0\frac{\partial p_2}{\partial \rho_2}\right|_{\rho_2=\rho_2^0, s=0} 
$ 
are the bulk moduli of the fluid and solid phases, respectively.

The deformation of the medium in case of small perturbations of the equilibrium 
solution can be described by any of the known equivalent strain tensors. In 
particular, one can use 
the Almansi strain tensor $\bm{\varepsilon} = (\bm{I} - \GG)/2$, where 
$\GG=\AA^T \AA$. 
For small perturbations of distortion $A_{ij}=\delta_{ij}+\delta 
A_{ij}$, 
the stress-strain relation \eqref{Thermod.forcesSF} takes the form 
\begin{align} \label{stress}
\sigma_{ij}=2\alpha_2 \rho_2 \csh^2 \left(\varepsilon_{ij}-\frac{\varepsilon_{11}+\varepsilon_{22}+\varepsilon_{22}}{3}\delta_{ij}\right),
\end{align}  
where $\varepsilon_{ij}=-(\delta A_{ij}+\delta A_{ji})/2$ are the components of the 
Almansi strain tensor $ \bm{\varepsilon} $. Thus, for small
perturbations we have
\begin{align} \label{stress1}
\delta \sigma_{ij}= \alpha_2^0 s_{ij} =
2\alpha_2^0 \rho_2^0 \csh^2 \left(\varepsilon_{ij}-\frac{\varepsilon_{11}+\varepsilon_{22}+\varepsilon_{22}}{3}\delta_{ij}\right).
\end{align}  
Small perturbation of $E_{\AA}=-\FF^T \bm{\sigma}$ reads as
\begin{align}
\delta E_{A_{ij}}=- \frac{\delta \sigma}{\rho^0}=-
2\alpha_2^0 \frac{\rho_2^0 \csh^2}{\rho^0} \left(\varepsilon_{ij}-\frac{\varepsilon_{11}+\varepsilon_{22}+\varepsilon_{22}}{3}\delta_{ij}\right).
\end{align} 

Now, omitting the notation $\delta$ for the perturbations, we obtain the 
following system:
\begin{subequations}\label{eqns.linear.complete0}
	\begin{eqnarray}
&&\frac{\partial v^i_1}{\partial t}+
\frac{1}{\rho_1^0}\frac{\partial p_1}{\partial x_i}- \frac{\alpha_2^0}{\rho^0}
\frac{\partial s_{ik}}{\partial x_k}
=-\frac{\alpha_2^0 \rho_2^0}{\rho^0}\frac{c_1^0c_2^0}{\theta_2}(v_1^i-v_2^i), 
 \\
&&\frac{\partial v^i_2}{\partial t}+
\frac{1}{\rho_2^0}\frac{\partial p_2}{\partial x_i}- \frac{\alpha_2^0}{\rho^0}
\frac{\partial s_{ik}}{\partial x_k}
=+\frac{\alpha_1^0 \rho_1^0}{\rho^0}\frac{c_1^0c_2^0}{\theta_2}(v_1^i-v_2^i), 
 \\
&&\frac{\partial \rho_1}{\partial t}+\rho_1^0 \frac{\partial v^k_1}{\partial 
x_k}=
\frac{\rho_1^0}{\alpha_1^0 \rho^0} \frac{p_2-p_1}{\theta_1},   \\
&&\frac{\partial \rho_2}{\partial t}+\rho_2^0 \frac{\partial v^k_2}{\partial 
x_k}=
-\frac{\rho_2^0}{\alpha_2^0 \rho^0} \frac{p_2-p_1}{\theta_1},   \\
&&\frac{\partial \alpha_1}{\partial t}= \frac{p_1-p_2}{\rho^0 \theta_1},  
 \\
&&\frac{\partial A_{i j}}{\partial t}+\delta_{im}\frac{\partial v^m}{\partial 
x_j}
=-\dfrac{ E_{A_{ik}} }{\theta} \label{eqns.linear.complete0.A} \\
&&\frac{\partial s}{\partial t}=0.  
\end{eqnarray}
\end{subequations}

One can see that entropy remains constant in time. That means that  
entropy 
has no influence on small perturbation wavefields, and the equation for 
entropy can be neglected. Also, because the shear stress $s_{ij}$ 
depends on the small strain tensor $\varepsilon_{ij}$,  the equation for 
distortion $\AA$ \eqref{eqns.linear.complete0.A} can be replaced by an
equation 
for $\bm{\varepsilon}$, which reads as
\begin{multline}
\frac{\partial \varepsilon_{ij}}{\partial t} - 
\frac{1}{2}\left(\frac{\partial v^i}{\partial x_j}+\frac{\partial v^j}{\partial x_i} \right)  
\\
= 
-\frac{2}{\theta}\alpha_2^0 \frac{\rho_2^0 \csh^2}{\rho^0} \left(\varepsilon_{ij}-\frac{\varepsilon_{11}+\varepsilon_{22}+\varepsilon_{22}}{3}\delta_{ij}\right).
\end{multline}

Finally, we obtain the following system: 
\begin{subequations}\label{eqns.linear.complete}
	\begin{eqnarray}
&&\frac{\partial v^i_1}{\partial t}+
\frac{1}{\rho_1^0}\frac{\partial p_1}{\partial x_i}- \frac{\alpha_2^0}{\rho^0}
\frac{\partial s_{ik}}{\partial x_k}
=-\frac{\alpha_2^0 \rho_2^0}{\rho^0}\frac{c_1^0c_2^0}{\theta_2}(v_1^i-v_2^i), 
 \\
&&\frac{\partial v^i_2}{\partial t}+
\frac{1}{\rho_2^0}\frac{\partial p_2}{\partial x_i}- \frac{\alpha_2^0}{\rho^0}
\frac{\partial s_{ik}}{\partial x_k}
=+\frac{\alpha_1^0 \rho_1^0}{\rho^0}\frac{c_1^0c_2^0}{\theta_2}(v_1^i-v_2^i), 
 \\
&&\frac{\partial \rho_1}{\partial t}+\rho_1^0 \frac{\partial v^k_1}{\partial 
x_k}=
\frac{\rho_1^0}{\alpha_1^0 \rho^0} \frac{p_2-p_1}{\theta_1},   \\
&&\frac{\partial \rho_2}{\partial t}+\rho_2^0 \frac{\partial v^k_2}{\partial 
x_k}=
-\frac{\rho_2^0}{\alpha_2^0 \rho^0} \frac{p_2-p_1}{\theta_1},   \\
&&\frac{\partial \alpha_1}{\partial t}= \frac{p_1-p_2}{\rho^0 \theta_1},  
 \\
&&\frac{\partial \varepsilon_{ij}}{\partial t} - 
\frac{1}{2}\left(\frac{\partial v^i}{\partial x_j}+\frac{\partial v^j}{\partial 
x_i} \right) = 
-\frac{2}{\theta}\alpha_2^0 \frac{\rho_2^0 \csh^2}{\rho^0}
\left(\varepsilon_{ij} \phantom{\frac12}\right. \notag\\
&&\left.  
\hspace{4.5cm}-\frac{\varepsilon_{11}+\varepsilon_{22}+\varepsilon_{22}}{3}\delta_{ij}\right).
  \end{eqnarray}
\end{subequations}

These equations form a basis for our study of small-amplitude waves in
a porous saturated medium at rest. 

Furthermore, we will impose a condition that will help further simplify the PDE 
system \eqref{eqns.linear.complete}. Our concern is to derive equations for 
waves which have 
wavelengths  much bigger than the characteristic size of the pores. 
This means that we assume instantaneous phase pressure equalizing. This 
is because the process is fully determined by pressure waves 
propagation and reflection at the pore boundaries and, thus, the 
characteristic time scale for reaching a pressure equilibrium is small.

The instantaneous relaxation of the phase pressures leads to governing 
equations obtained in the relaxation limit of \eqref{eqns.linear.complete} as 
$\theta_1 \rightarrow 0$.
In this case the resulting system is the system 
\eqref{eqns.linear.complete} in which the equation for $\alpha_1$ is replaced 
by an algebraic equation: $p_1=p_2=P$. Then, since the phase pressures are 
equal,  we have from \eqref {pressure.linear} that
$$\frac{K_1}{\rho^0_1} \rho_1=\frac{K_2}{\rho^0_2} \rho_2$$  
and, hence, the two equations for the phase densities $\rho_1,\rho_2$ can be 
replaced by 
a single equation for the pressure $P$:
\begin{align}\label{pressure.equation}
\frac{\partial P}{\partial t} +\frac{\alpha_1^0}{\alpha_1^0 K_1^{-1}+\alpha_2^0 K_2^{-1}}
\frac{\partial v_1^k}{\partial x_k}+
\frac{\alpha_2^0}{\alpha_1^0 K_1^{-1}+\alpha_2^0 K_2^{-1}}
\frac{\partial v_2^k}{\partial x_k} =0.
\end{align}

Now, taking into account \eqref{stress}, \eqref{pressure.equation} and $v_1^k=v^k+c_2^0w^k$, 
$v_2^k=v^k-c_1^0w^k$,
we obtain the resulting system written in terms of the mixture velocities, relative velocities, pressure, and shear stress:
\begin{subequations}\label{stress.velocity}
	\begin{eqnarray} 
&& \rho^0 \frac{\partial v^i}{\partial t}+\frac{\partial P}{\partial x_i}-
\alpha^0_2 \frac{\partial s_{ik}}{\partial x_k} = 0,  \\
&& \frac{\partial w^k}{\partial t}+ \left(\frac{1}{\rho_1^0} - 
\frac{1}{\rho_2^0}\right)\frac{\partial P}{\partial x_i}=
-\frac{c_1^0c_2^0}{\theta_2}w^k,  \label{stress.velocity.w}\\
&& \frac{\partial P}{\partial t} +K\frac{\partial v^k}{\partial x_k}+
\frac{\alpha_1^0\alpha_2^0}{\rho^0}\left(\rho^0_2-\rho^0_1 \right)
K\frac{\partial w^k}{\partial x_k} =0,   \\   \label{sij}
&&\frac{\partial s_{ik}}{\partial t} - 
\mu\left(\frac{\partial v^i}{\partial x_k} +\frac{\partial v^k}{\partial x_i}-
\frac{2}{3}\delta_{ik}\frac{\partial v^j}{\partial x_j} \right) = 
- \alpha_2^0 \frac{s_{ik}}{\tau}  
\end{eqnarray}
\end{subequations}
where $K=\left(\alpha_1^0 K_1^{-1}+\alpha_2^0 K_2^{-1}\right)^{-1}$, 
$\mu=\rho_2^0 \csh^2$, and $\tau $ 
is the shear stress relaxation time.

Thus, we have derived the PDE system \eqref{stress.velocity} for small 
amplitude wave propagation in a saturated porous medium at rest. In this 
system, two dissipation mechanisms are present: (i) friction of the
fluid on the pore walls and (ii) shear stress relaxation of the 
saturated porous medium. In the following section, some properties of the 
wavefields governed by \eqref{stress.velocity} will be studied.

\section{Comparison of SHTC and Biot models}\label{Comparison}

\subsection{Theoretical comparison}

The most comprehensive comparative study of Biot's model and models developed 
by a classical two-phase approach based on continuum thermodynamics has  
been made by Wilmanski \cite{Wilmanski2006}.
The conclusion of this paper is that the two-phase model contains all 
information in Biot's model about the features of wave propagation in a saturated porous 
medium, and, in particular, it predicts the slow pressure 
waves and gives a qualitatively correct description of the dependence of the phase 
velocities on frequency.

The governing equations of the SHTC model proposed in the present paper differ from 
those obtained by the classical two-phase solid-fluid approach, but we will see 
that qualitatively the properties of wavefields are the same as those in 
Biot's model.
We emphasise that the above-proposed SHTC model can be used for describing the finite deformations in the medium by taking into account changes in porosity under stress variations and inelastic deformations. 
It should be also emphasized that the consideration of finite deformations 
imposes certain restrictions on the definition of deformations of the constituents. 
For example, the density of the mixture is an additive parameter which can be 
defined as the sum of partial densities, $ \rho=\alpha_1 \rho_1 + \alpha_2 
\rho_2$. On the other hand, the decomposition of arbitrary finite deformation 
into 
deformations of the solid and fluid phases is questionable, because this could 
imply the existence of two relaxed reference frames and two sets of Lagrangian 
coordinates. That is why we consider distortion $\AA$ as a measure of 
deformation of the whole medium and do not consider deformation of the skeleton 
separately. In this context, we recall that our unified continuum model 
for fluids and solids \cite{HPR2016,DPRZ2016,HYP2016} also relies on a 
deformation-based rather than a strain-rate-based description of fluid flows.

The stress-strain relation in the SHTC model fully depends on the definition of 
the thermodynamic potential $ E $ \eqref{energy.SF}. The total stress which is 
presented in the total momentum equation \eqref{eqn.momentumPV} reads as 
\begin{align}\label{OurTotal.stress}
T_{ik}=\alpha_2 (\sigma_{ik}-p_2)-\alpha_1 p_1,
\end{align}
where $\sigma_{ik}$ is the deviatoric trace-free stress tensor ($\tr 
(\bm{\sigma})=0$). 
In Biot's model, the total stress is defined as 
\begin{align}\label{BiotTotal.stress}
T_{ik}=\sigma^{\textrm ef}_{ik}-n P\delta_{ik},
\end{align}
where $\sigma^{\textrm ef}_{ik}$ is the so-called effective stress, $P$ is the pore 
pressure, and $n$ is the Biot coefficient (see, for example, 
\cite{Merxhani2016}). 
Thus, if we take $\sigma^{\textrm ef}_{ik}=\alpha_2\sigma_{ik}$ in our SHTC 
formulation \eqref{OurTotal.stress}, we have
\begin{equation}
T_{ik}=\sigma^{\textrm ef}_{ik}-\alpha_1 p_1- \alpha_2 p_2,
\end{equation}
and in the case of instantaneous pressure relaxation with $p_1=p_2=P$, we 
obtain Biot's formula \eqref{BiotTotal.stress} with 
$n=1$.

Furthermore, note that there is a possibility to specify the energy 
potential in such a way that 
formula \eqref{BiotTotal.stress} 
holds also in the SHTC model.
To do this we define the volumetric part of internal energy 
as
\begin{align*} 
\rho E_1=n c_1 \alpha_1 \rho_1e_1(\rho_1,s)+\alpha_2 \rho_2 e_2(\rho_2,s) 
\end{align*}
instead of \eqref{energy.mix},
and take the pressure relaxation term in the equation for volume fraction as
$\frac{ n p_1-p_2}{\rho \theta_1}$. In this paper we do not discuss this 
possibility in detail. 


As to the momentum balance for the fluid constituent in the SHTC 
model, there is 
an additional force to induce fluid flow, which is connected with the shear 
stress gradient and  not presented in Biot's model. 
Specifically, with the notation $v^i_1=\vf$, $v^i_2=\vs$ for the velocities, 
$\rho^0_1=\rhof$, $\rho^0_2=\rhos$ for the densities,  and
$c_1=\cf$, $c_2=\cs$ for the mass fractions, 
equation \eqref{stress.velocity.w} for the relative velocity can be replaced by 
the following equivalent equation for the momentum of the fluid:
\begin{eqnarray}\label{vs.SHTC}
&& \rhof\frac{\partial \vf^i}{\partial t}+ \frac{\partial P}{\partial 
	x_i}-\frac{\rhof}{\rho_0}
\frac{\partial \sigma_{ik}}{\partial x_k}=
-\frac{\rhof\cs^2}{\theta_2}(\vf^i-\vs^i),
\end{eqnarray}
where $\rho^0=\alphaf\rhof+\alphas\rhos$ is the mixture density, 
$P$ is the fluid pressure, $\sigma_{ik}=\alphas s_{ik}$ is 
the shear stress, and $\cf,\cs$ are the mass fractions of the fluid and solid constituents, respectively. The 
shear stress gradient in the left hand-side of \eqref{vs.SHTC} is a term 
which is present in the 
SHTC equations and absent in Biot's model, e.g. see equation 
\eqref{Biot.model.vs} below.  In fact, from a physical standpoint, taking 
into 
account the shear stress in the balance equation for the fluid momentum seems 
to be rather natural, because the motion of the fluid may be caused not only by 
the pressure gradient, but also by tangential deformation of the mixture  
element.

Overall, the equations for the pressure and shear stress in the SHTC and 
Biot's models are quite different and cannot be transformed one into another,  
because the stress-strain relationships and elastic moduli used 
in these models are different. Nevertheless, it will be shown below that the features of 
wavefields are qualitatively similar in both models, and in some cases they are 
quantitatively close  by a corresponding choice of the material parameters. In the 
following section, this statement will be discussed in a more quantitative manner via a plane wave 
analysis.


\subsection{Analysis of wavefields in saturated porous 
 	media}\label{sec.waves}
 
The 3D system of governing equations for small-amplitude wave propagation  
\eqref{stress.velocity} is quite complex, but an informative analysis can be 
done for 1D plane waves. In this section, we neglect the stress 
relaxation term in the equation for the shear stress tensor of the SHTC model, as it 
is absent in  
Biot's model.

%

\subsubsection{1D Biot's theory}

The one-dimensional first-order form of the governing equations of the 
so-called ``low-frequency 
limit'' in Biot's theory for the vector of state variables $ (v_{\textrm s}, q, 
\sigma,p) 
$ can be formulated as follows 
\cite{Carcione2010, Masson2006}:
\begin{subequations}\label{eqn.Biot.model}
	\begin{align}
		& \rho_{\textrm f} \frac{\pd v_{\textrm s}}{\pd t} + \rho_{\textrm f} \frac{\tort}{\phi} 
		\frac{\pd 
		q}{\pd t} + 
		\frac{\pd p}{\pd x} = - 
		\frac{\eta}{\kappa} q , \label{Biot.model.vs}\\[2mm]
		& \rho \frac{\pd v_{\textrm s}}{\pd t} + \rho_{\textrm f} \frac{\pd q}{\pd t} - \frac{\pd 
		\sigma}{\pd x} = 
		0,\\[2mm]
		& \frac{\pd \sigma}{\pd t} - (\lambda_{\textrm u} + 2 \mu_{\textrm u}) \frac{\pd v_{\textrm 
		s}}{\pd x} 
		- \alpha M 
		\frac{\pd 
		q}{\pd x} = 0, \\[2mm]
		& \frac{\pd p}{\pd t} + \alpha M \frac{\pd v_{\textrm s}}{\pd x} + M \frac{\pd q}{\pd x} = 
		0,
	\end{align}
where $ \phi $ is the porosity, $ v_{\textrm s} $ and $ q = \phi 
(v_{\textrm f} - v_{\textrm s}) $ are the solid 
and fluid (relative to 
the 
solid) particle velocities, $ \sigma $ and $ p $ are the bulk 
stress and fluid pressure, 
respectively. Additionally, $ \rho_{\textrm f} $ is the fluid density and $ \rho = 
(1-\phi)\rho_{\textrm s} 
+\phi\rho_{\textrm f} $ is the total (mixture) density. Furthermore, the material parameters are 
defined as follows:
\begin{align}\label{eqn.Biot.param1}
&\lambda_{\textrm u} = K_{\textrm u} - \frac{2}{3}\mu_{\textrm u} = \Km + \alpha^2 M - 
\frac{2}{3}\mu_{\textrm u}, \notag \\
&  M = B K_{\textrm u}/\alpha, \ \ \alpha = 1- \Km/K_{\textrm u}
\end{align}
\begin{equation}\label{eqn.Biot.param2}
\Ku = \frac{\Km}{1 - \alpha B}, \qquad B = \frac{1/\Km - 1/\Ks}{1/\Km - 1/\Ks + \phi(1/\Kf - 
1/\Ks)},
\end{equation}
\end{subequations}
where the subscript ``u'' denotes the quantities characterizing the \textbf{u}ndrained solid matrix, while 
the subscript ``m'' denotes the quantities characterizing the dry \textbf{m}atrix. Thus, 
$ \lambdau $, $ \Ku $, and $ \mu_{\textrm u} $ are the undrained Lam\'e parameter, bulk modulus, 
and 
the shear modulus of the undrained matrix, respectively, while $ \Km $ is the bulk modulus of the dry matrix. 
Additionally, $ \Ks $ and $ \Kf $ are the bulk moduli of grains and fluid, respectively, while $ B $ is the 
so-called Biot parameter, and $ \tort $ is the tortuosity. Also, it is implied that the shear 
modulus of the undrained matrix and that of the dry one coincide, $ \muu=\mu_{\textrm m} $.

\subsubsection{Characteristic speeds}\label{sec.char.speeds}

Let us assume only elastic deformation of the porous material and consider 
the linearized SHTC system 
\eqref{stress.velocity} in one dimension ($ x = x_1 $) for the vector of state 
variables
\begin{equation}
\QQ = (v_1,v_2,v_3,w_1,w_2,w_3,p,s_{11},s_{21},s_{31}).
\end{equation}
System \eqref{stress.velocity} can be written in  matrix form as follows:
\begin{equation}\label{eqn.1Dfull}
\QQ_t + \mathbb{A} \QQ_x = \SS,
\end{equation}
where 
\begin{align}\label{eqn.1Dfull.mat}
&\mathbb{A} = \notag\\
&
\setlength{\arraycolsep}{1.75pt} 
\left( \begin{array}{cccccccccc}
	     0      &  0   &  0   & 0  & 0 & 0 & \rho^{-1} & -\alpha_2 \rho^{-1} &          0          
	     &          0          \\
	     0      &  0   &  0   & 0  & 0 & 0 &     0     &          0          & -\alpha_2 \rho^{-1} 
	     &          0          \\
	     0      &  0   &  0   & 0  & 0 & 0 &     0     &          0          &          0          
	     & -\alpha_2 \rho^{-1} \\
	     0      &  0   &  0   & 0  & 0 & 0 &     R     &          0          &          0          
	     &          0          \\
	     0      &  0   &  0   & 0  & 0 & 0 &     0     &          0          &          0          
	     &          0          \\
	     0      &  0   &  0   & 0  & 0 & 0 &     0     &          0          &          0          
	     &          0          \\
	     K      &  0   &  0   & K' & 0 & 0 &     0     &          0          &          0          
	     &          0          \\
	-\frac43\mu &  0   &  0   & 0  & 0 & 0 &     0     &          0          &          0          
	&          0          \\
	     0      & -\mu &  0   & 0  & 0 & 0 &     0     &          0          &          0          
	     &          0          \\
	     0      &  0   & -\mu & 0  & 0 & 0 &     0     &          0          &          0          
	     &          0
\end{array} 
\right), 
\notag\\
&\SS = \left(\begin{array}{c}
0 \\ 
0 \\ 
0 \\ 
-\frac{1}{\theta_2'} w^1 \\ 
-\frac{1}{\theta_2'} w^2 \\ 
-\frac{1}{\theta_2'} w^3 \\ 
0 \\ 
0 \\ 
0 \\ 
0
\end{array} \right) 
\end{align}
with $ R = 1/\rho_1 - 1/\rho_2 $, $ K' =
\frac{\alpha_1^0\alpha_2^0}{\rho^0}\left(\rho^0_2-\rho^0_1
\right)
K $,  and and $ \theta_2' = \theta_2/(c_1^0 c_2^0) $. The characteristic
speeds  $ \lambda_i $ of system 
\eqref{eqn.1Dfull} (the 
eigenvalues of $ \mathbb{A} $) are the roots of the characteristic polynomial $ 
\det(\mathbb{A} - \lambda \mathbb{I}) = 0 $. These roots are 
given by the formulas
\begin{subequations}\label{speeds.SHTC}
	\begin{align}\label{SHTC.charar.pressure}
&\lambda^2 = 0, \quad \lambda^2 = \frac{\alpha_2 \mu}{\rho},\notag\\ 
&\lambda^2 = 
\frac{X+Y \pm 
\sqrt{(X+Y)^2 - 4(X - Z)Y}}{2},
\end{align}
	\begin{equation}
		X = K/\rho + R K', 
		\qquad Y = \frac{4}{3} \alpha_2 \mu/\rho , 
		\qquad
		Z = K/\rho .
	\end{equation}
\end{subequations}
Formula \eqref{SHTC.charar.pressure}$ _3 $ provides the so-called fast $ C_{\textrm 
fast} $ and slow $ C_{\textrm slow} $ characteristic speeds corresponding to  
longitudinal waves (fast and slow P-waves).

For Biot's model \eqref{eqn.Biot.model} we have 
\begin{equation}\label{eqn.Mat.Biot}
\mathbb{A} = \left(
\begin{array}{cccc}
0 & 0 & \frac{\phi}{z} & \frac{\tort}{z} \\ 
0 & 0 & \frac{-\phi\rho}{\rhof z} & \frac{-\phi}{z} \\ 
\alpha M & M & 0 & 0 \\ 
 -\lambdau - 2\mu_{\textrm u} & -\alpha M & 0 & 0
\end{array} 
\right), \quad z = \phi\rhof - \rho\tort,
\end{equation}
and the characteristic polynomial is
\begin{subequations}
	\begin{equation}
		\det(\mathbb{A} - \lambda \mathbb{I}) = \lambda^4 + a_2 \lambda^2 + a_0 = 0,
	\end{equation}
	\begin{align}
		&	a_2 = \frac{\rhof\tort (\lambdau + 2 \muu) + \phi M (\rho - 2 \alpha 
		\rhof)}{\rhof(\phi\rhof - \rho \tort)}, \notag\\
		& 
		a_0 = \frac{\phi M(\alpha^2 M - \lambdau - 2 \muu)}{\rhof(\phi\rhof - 
		\rho \tort)},
	\end{align}
while the characteristic speeds are given by the formula
\begin{equation}
C_{\textrm fast} = \frac{-a_2 + \sqrt{a_2^2 - 4 a_0}}{2},
\qquad
C_{\textrm slow} = \frac{-a_2 - \sqrt{a_2^2 - 4 a_0}}{2},
\end{equation}
\end{subequations}
which provides the fast and slow  characteristic speeds corresponding to  P-waves 
in Biot's model.

\begin{figure}[!htbp]
	\begin{center}
		\includegraphics[draft=false,
		width=0.75\linewidth]{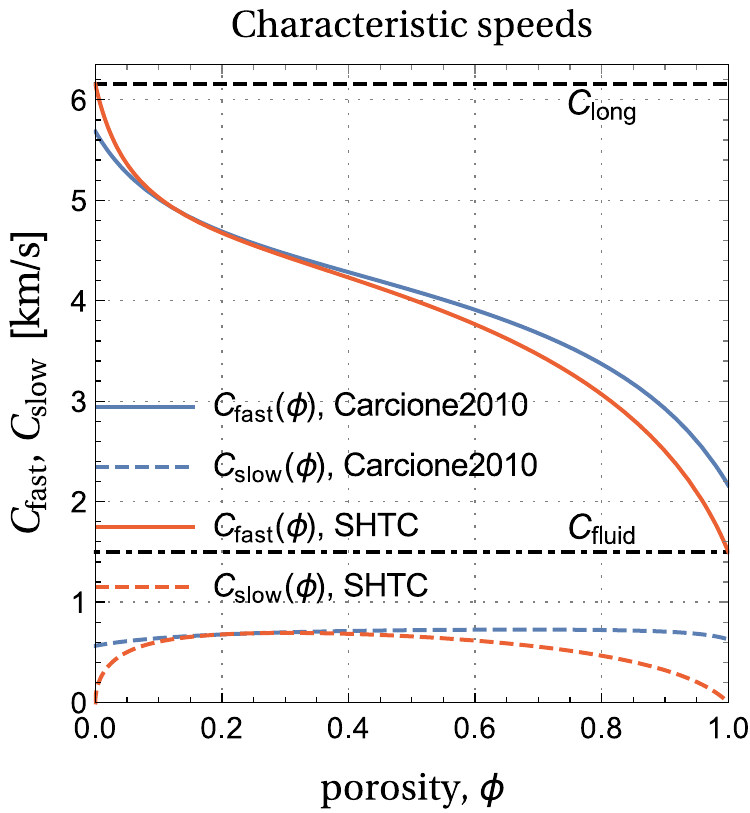}
		\caption{Comparison of the characteristic speeds for the Biot and SHTC 
		models for the material parameters given in 
		Table\,\ref{tab:parameters}. Biot's model is taken in the version developed by 
		Carcione et al \cite{Carcione2010}.  
				}  
		\label{fig:diff.grain.modulus}
\end{center}
\end{figure}

In Fig\,\ref{fig:diff.grain.modulus}, for the two models, we compare the 
characteristic speeds $ C_{\textrm fast}(\phi) $ and $ C_{\textrm 
slow}(\phi) $ of the fast 
and slow modes. Some other material parameters 
are given in Table\,\ref{tab:parameters}. Note that, in order to make the 
curves closer to each other, the bulk moduli of the solid
phase in the two models are taken slightly different. 

In the limit cases $ \phi \to 0 $ (pure solid) and $ \phi \to 1 $ (pure fluid) the behavior of 
the characteristic speeds in the SHTC model better corresponds to that 
intuitively 
expected, i.e. the characteristic speeds of the mixture coincide with the 
characteristic speeds of the pure phases (horizontal dashed and dashed-dotted 
lines), while the slow mode vanishes. In 
contrast, in Biot's model the slow mode is present even in the pure fluid or 
pure solid cases. 

\begin{table}[t]
	\begin{center}
		\setlength{\tabcolsep}{1.75pt}
		\begin{tabular}{llll}
			\hline
			Parameters\hspace{1cm}                   & Biot\hspace{2cm} & 
			SHTC          & 
			Unit                           \\ \hline
			\textbf{Grain:}		& & &\\[1mm]
			\rowcolor[HTML]{EFEFEF}
			$ \rhos = \rho_2 $      & 2500   &        2500             & kg/m$ 
			^3 
			$                      \\[1mm]
			$ \Cs  $     &  4000       &   4332        & 
			m/s                              \\[1mm]
			\rowcolor[HTML]{EFEFEF}
			$ \Csh  $  &  -       &   3787        & 
			m/s                              \\[1mm]
			$ \Ks = K_2 = \rho_2 \Cs^2 $  &   40       &   46.9        & 
			GPa      \\[1mm]
			\rowcolor[HTML]{EFEFEF}
			$ \mus = \mu = \rho_2 \Csh^2 $&   -       &  35.85        & GPa 
			\\[1mm]
			\textbf{Matrix, $\phi =0.2$:}		& & &\\[1mm]
			\rowcolor[HTML]{EFEFEF} 			
			$ \Km $     & $ 9.48 $       &    -     & GPa                      
			\\[1mm]
			$ \mu_{\textrm m} $     & $ 24.5 $       &    -     & 
			GPa                      \\[1mm]
			\rowcolor[HTML]{EFEFEF} 
			$ \tort $     & $ 3.75 $           &   -       & 
			-                            \\[1mm]
			$ \kappa $ & $ 1\cdot10^{-13} $      &     -     & m$ ^2 
			$                	\\[1mm]
			\rowcolor[HTML]{EFEFEF}
			$ \theta_2 $       & - &  $ 3.36\cdot10^{-7}$  & s  \\[1mm]
			\textbf{Fluid:}		& & &\\[1mm]
			\rowcolor[HTML]{EFEFEF}
			$ \rhof=\rho_1 $ & 1040        &     1040       & kg/m$ ^3 
			$              \\[1mm]
			$ \Cf $ & 1500        &     1500       & 
			m/s                              \\[1mm]
			\rowcolor[HTML]{EFEFEF}
			$ \Kf =K_1=\rhof \Cf^2$     & 2.34          &   2.34       & 
			GPa                   
			\\[1mm]
			$ \eta $  & $ 10^{-3} $&    -      & Pa$ \cdot $ s             
			\\[1mm]
			\hline
		\end{tabular}
		\caption{ Material parameters for the Biot and SHTC models. Here 
		$C_{\textrm s}$ and $C_{\textrm f}$ are the bulk sound speeds in the solid and 
		fluid, respectively.}
		\label{tab:parameters}
	\end{center}
\end{table}

\subsubsection{Dispersion relations and sound speeds}

In this section, we will perform a plane wave analysis of the SHTC and Biot 
models. Moreover, we will 
consider only longitudinal waves. In this case the vector of SHTC state 
variables reduces to 
\begin{equation}
\QQ = (v_1,w_1,p,s_{11}).
\end{equation}
By linearizing the right hand-side of system \eqref{eqn.1Dfull} in a 
neighborhood of $ \QQ = \QQ_0 + \qq $, where $ \qq 
$ is a small perturbation of an equilibrium state $ \QQ_0 = (0,0,p_0,0) $, we have
\begin{equation}\label{eqn.perturb}
\qq_t + \mathbb{A}\qq_x = \mathbb{S}\qq,
\end{equation}
where $ \mathbb{A} $ and $ \mathbb{S} = \pd\SS/\pd\QQ $ are matrices taken 
at the equilibrium state $ \QQ_0 $:
\begin{align}\label{eqn.1d.matrices}
&\mathbb{A} = \left( \begin{array}{cccc}
0 & 0 & \rho^{-1} & \alpha_{2}\rho^{-1} \\ 
0 & 0 & R & 0 \\ 
K & K' & 0 & 0 \\ 
-\frac43\mu & 0 & 0 & 0
\end{array} \right), \notag\\
&
\mathbb{S} = \left( \begin{array}{cccc}
0 & 0 & 0 & 0 \\ 
0 & -\frac{1}{\theta_2'} & 0 & 0 \\ 
0 & 0 & 0 & 0 \\ 
0 & 0 & 0 & 0
\end{array} \right).
\end{align}

We seek a solution that has the form
\begin{equation}\label{eqn.qq}
\qq = \tilde{\qq} e^{\text{i}(\omega t - k x)},
\end{equation}
which represents a plane harmonic wave of real frequency $ \omega $ and complex wave number $ k $ 
propagating in the direction $ x $, and $ \tilde{\qq} = const $ is a constant 
vector of amplitudes. 
By substituting \eqref{eqn.qq} in \eqref{eqn.perturb}, we arrive at a homogeneous linear system for 
$ \tilde{\qq} $ \cite{Ruggeri1992,Ruggeri2015}:
\begin{equation}\label{eqn.lin.dispers}
\left(\mathbb{I} - \frac{k}{\omega} \mathbb{A} + \frac{\text{i}}{\omega} \mathbb{S} \right) 
\tilde{\qq} = 0
\end{equation}
where $ \mathbb{I} $ is the identity matrix of the same size as $ \mathbb{A} $ 
and $ \mathbb{S} $. From \eqref{eqn.lin.dispers} we have the following dispersion relation for 
\eqref{eqn.1Dfull}:
\begin{equation}\label{eqn.dispers.rel}
\det\left(\mathbb{I} - \frac{k}{\omega} \mathbb{A} + \frac{\text{i}}{\omega} \mathbb{S} \right) = 0.
\end{equation}
The phase velocity $ V_{\textrm ph} $ (sound speed) and the attenuation factor $ a 
$ are then given by
\begin{equation}
V_{\textrm ph} = \frac{\omega}{{\textrm Re}(k)}, \qquad a = -{\textrm Im}(k).
\end{equation}
In addition, it is convenient to use the attenuation per wavelength 
\cite{Ruggeri2015}
\begin{equation}\label{atten.wavelength}
a_\lambda = a \lambda = \frac{2 \pi V_{\textrm ph} a }{\omega} = -2 \pi \frac{{\textrm 
Im}(k)}{{\textrm Re}(k)},
\end{equation}
where $ \lambda $ is the wavelength.

There are four solutions to \eqref{eqn.dispers.rel} with matrices 
\eqref{eqn.1d.matrices}, which are 
	\begin{equation}
	{\scriptstyle   
	k(\omega) = \pm \omega \sqrt{\frac{X+Y - {\textrm i}(Y+Z)/\Omega\pm \sqrt{4 Y 
	(X-Z)({\textrm i}/\Omega-1) + 
	({\textrm i}(Y+Z)/\Omega - (X + Y))^2}}{2(X-Z)Y}},
	}
	\end{equation}
where $ X, Y $, and $ Z $ are defined in \eqref{speeds.SHTC}, and $ \Omega = 
\omega \theta_2' $ is the non-dimensional frequency.

Applying the same analysis to the Biot equations \eqref{eqn.Biot.model}, we 
find that the dispersion relation is a bi-quadratic equation 
\begin{subequations}
	\begin{equation}\label{Biot.disp}
		\lambda^4 + a_2 \lambda^2 + a_0 = 0,
	\end{equation}
	\begin{align}\label{a2.Biot}
	&	a_2 = \omega  \left(-\frac{\omega  (\tort \rhof 
		(\lambdau +2 \muu )+M \phi  (\rho -2 \alpha  \rhof))}{\phi 
		M \left(\lambdau +2 \muu - \alpha ^2 M\right)} \right .\notag \\
	& \left. +\frac{{\textrm i} \eta  
		(\lambdau +2 \muu )}{\kappa M \left(\lambdau +2 \muu - \alpha ^2 
		M\right)}\right),
	\end{align}
	\begin{equation}\label{a0.Biot}
		a_0 = \omega ^3 \left(\frac{\rhof \omega  (\tort \rho -\rhof \phi 
		)}{\phi M  \left(\lambdau +2 \muu - \alpha ^2 M\right)}-\frac{{\textrm i} 
		\eta  
		\rho }{\kappa M \left(\lambdau +2 \muu - \alpha ^2 M\right)}\right).
	\end{equation}
\end{subequations}

\begin{figure*}[t]
	\begin{center}
			\includegraphics[draft=false,trim=0 0 0 0,clip, 
			scale=0.5]{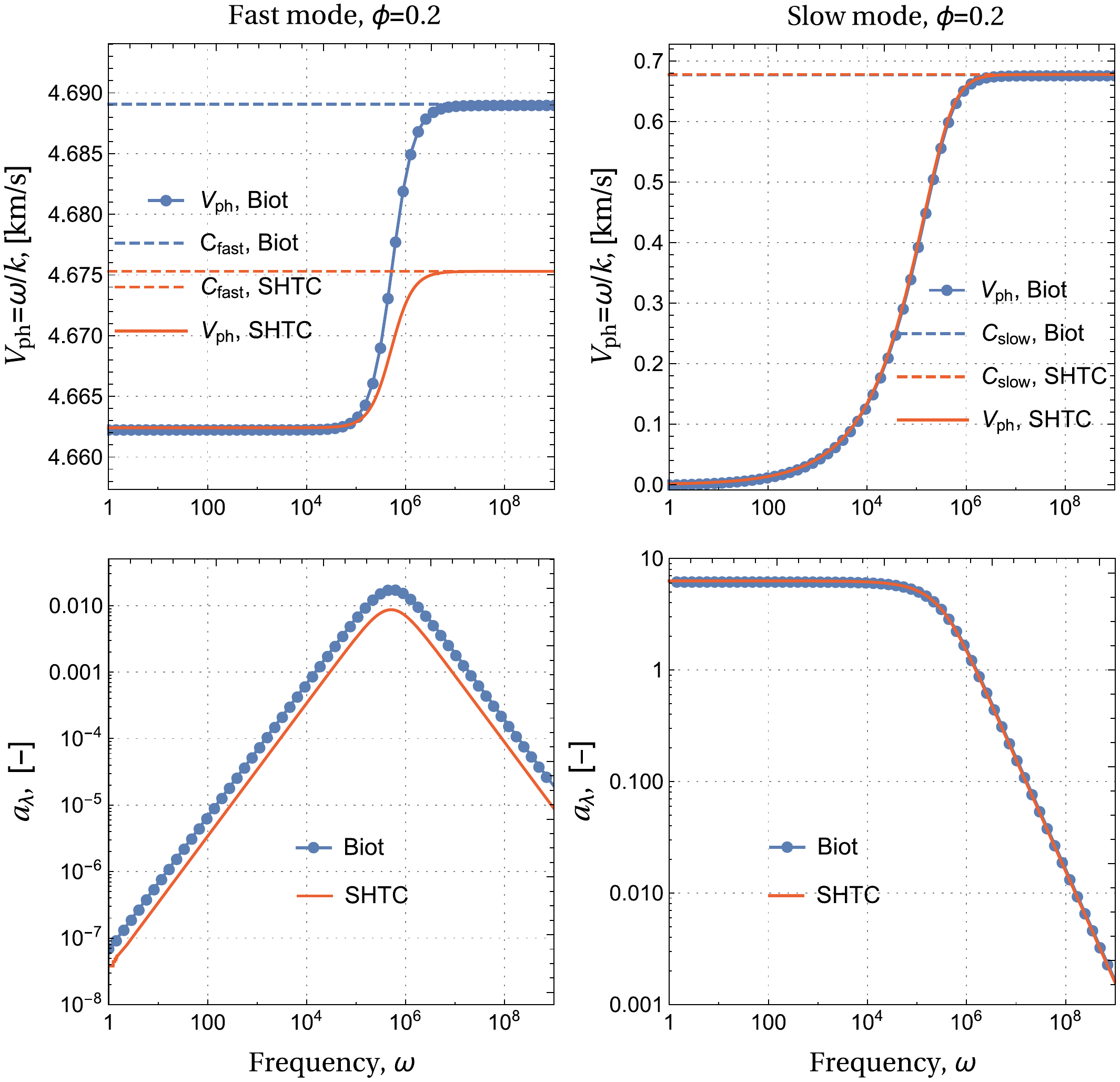}
		\caption{Phase velocities (first row) and attenuation factors per 
			wavelength (second row) for the fast (first column) and slow 
			(second column) modes of the SHTC and Biot models. The material 
			parameters are given in Table\,\ref{tab:parameters}.  
		}  
		\label{fig:Vph.Attenuation}
	\end{center}
\end{figure*}

The phase velocities $ V_{\textrm ph}(\omega) $ and attenuation 
factor per wavelength $ a_{\lambda}(\omega) $ of the fast and slow sound waves 
for both models are shown in Fig.\,\ref{fig:Vph.Attenuation}. Note that there is a difference in the high-frequency limit of the fast modes, 
which is, in fact, due to the 
difference between the characteristic speeds $ C_{\textrm fast} $ and $ C_{\textrm slow} $ 
depicted in 
Fig.\,\ref{fig:diff.grain.modulus}. To explain this, recall that the 
characteristic 
speeds (the eigenvalues of the homogeneous hyperbolic system) are the 
high-frequency limits ($ \omega \to \infty $) of the sound speeds (the 
eigenvalues of the non-homogeneous system \eqref{eqn.perturb}) 
\cite{Ruggeri1992,Ruggeri2015,DPRZ2016}.
The slow mode dispersion curves of both models are almost 
indistinguishable, see 
Fig.\,\ref{fig:Vph.Attenuation}, the second column.

Another conclusion is that there is a big difference ($ \sim 10^3 $) in the attenuation 
factors $ a_\lambda $ of the fast and slow modes for both models in the low 
frequency region, see Fig.\,\ref{fig:Vph.Attenuation}, the second row.

%

\section{Numerical test problems for small-amplitude wave 
propagation}\label{sec.numerics}

\subsection{Finite difference implementation}
To discretize the governing equations, the velocity-stress formulation, which was
proposed for elastic-wave equations in  \cite{Levander1988,Virieux1986}, is used. 
The use of a numerical scheme on staggered grids is quite natural due to 
the fact that the equations form a symmetric first order system of evolution 
equations for the  mixture components, relative velocities, pressure, and 
shear stress.

Following the notation of \cite{Virieux1986} and \cite{Graves1996}, we 
introduce a time-space grid with integer nodes $t^{n}=n\Delta t$, 
$x_{i}=i\Delta 
x$, $y_{j}=j\Delta y$  and half-integer nodes $t^{n+1/2}=(n+1/2)\Delta t$, 
$x_{i+1/2}=(i+1/2)\Delta x$, $y_{j+1/2}=(j+1/2)\Delta y$ , where  $\Delta t$, 
$\Delta x$, and $\Delta y$ denote the grid step sizes for the temporal and spatial 
axes.

 For a discrete function $f_{i,j}^{n}=f(t^{n},x_{i},y_{j})$ defined at the grid 
 nodes, let us introduce second-order  centered finite difference 
 operators:
\begin{align}
    & D_t[f]_{i, j}^{n} = \frac{(f)_{i, j}^{n+1/2} - (f)_{i, j}^{n-1/2}}{\Delta t}, \notag\\
    & A_t[f]_{i, j}^{n} = \frac{(f)_{i, j}^{n+1/2} + (f)_{i ,j}^{n-1/2}}{2},
    \label{Dt}
\end{align}
\begin{align}
   & D_x[f]_{i, j}^n = \frac{(f)_{i+1/2, j}^{n} - (f)_{i-1/2, j}^{n}}{\Delta x}, \notag \\ 
   & D_y[f]_{i, j}^n = \frac{(f)_{i, j+1/2}^{n} - (f)_{i, j-1/2}^{n}}{\Delta y}.
    \label{Dx}
\end{align}

The medium parameters are constant within each grid cell 
$[x_{i-1/2}, x_{i+1/2}] \times[ y_{j-1/2}, y_{j+1/2}]$, and they may have 
discontinuities aligned with grid lines. This condition provides  
second-order convergence even for discontinuous medium parameters 
\cite{Moszo2002}.

The wavefield components are defined at different time-space grid nodes. The 
components of the mixture and the relative velocities are defined as $(v_x)_{i+1/2, 
j}^{n}$, $(v_y)_{i, j+1/2}^{n}$, $(w_x)_{i+1/2, j}^{n}$, $(w_y)_{i, 
j+1/2}^{n}$, the pressure and the normal components of the deviatoric stress as 
$(p)_{i,j}^{n+1/2}$,  $(s_{xx})_{i,j}^{n+1/2}$, $(s_{yy})_{i,j}^{n+1/2}$, and 
the shear stress as $(s_{xy})_{i+1/2, j+1/2}^{n+1/2}$.

To construct the finite difference scheme, we use a finite volume 
approximation  or balance law technique \cite{Samarskii2001}. In this case the 
discrete form 
of equations \eqref{stress.velocity} reads as
\begin{subequations}
\begin{align}\label{eq:v1}
&
D_t[v_x]_{i+1/2 ,j}^{n-1/2} =- \left \langle 1/\rho^{0} \right \rangle_{i+1/2,j}D_x[P]_{i+1/2 
,j}^{n-1/2} \notag \\
& + \left \langle\alpha_{2}^{0}/ \rho^{0} \right \rangle_{i+1/2,j} \left( D_x[s_{xx}]_{i+1/2, 
j}^{n-1/2} + D_y[s_{xy}]_{i+1/2, j}^{n-1/2} \right), 
\end{align}
\begin{align}\label{eq:v2}
&D_t[v_y]_{i ,j+1/2}^{n-1/2} =- \left \langle 1/\rho^{0} \right \rangle_{i,j+1/2}D_y[P]_{i 
,j+1/2}^{n-1/2} \notag \\[2mm]
& + \left \langle\alpha_{2}^{0}/ \rho^{0} \right \rangle_{i,j+1/2} \left( D_x[s_{xy}]_{i, 
j+1/2}^{n-1/2} + D_y[s_{yy}]_{i, j+1/2}^{n-1/2} \right), 
\end{align}
\begin{align}\label{eq:w1}
& D_t[w_x]_{i+1/2 ,j}^{n-1/2} =- \left \langle 1/\rho_{1}^{0}-1/\rho_{2}^{0} \right 
\rangle_{i+1/2,j}D_x[P]_{i+1/2 ,j}^{n-1/2} \notag \\[2mm]
& -\left \langle c_{1}^{0}c_{2}^{0}/\theta_{2} \right \rangle_{i+1/2,j}A_t[w_x]_{i+1/2 ,j}^{n-1/2},
\end{align}
\begin{align}\label{eq:w2}
&
D_t[w_y]_{i ,j+1/2}^{n-1/2} =- \left \langle 1/\rho_{1}^{0}-1/\rho_{2}^{0} \right 
\rangle_{i,j+1/2}D_y[P]_{i ,j+1/2}^{n-1/2} \notag \\[2mm]
& -\left \langle c_{1}^{0}c_{2}^{0}/\theta_{2} \right \rangle_{i,j+1/2}A_t[w_y]_{i ,j+1/2}^{n-1/2} 
,
\end{align}
\begin{align}\label{eq:s11}
&D_t[s_{xx}]_{i ,j}^{n} = (\mu)_{i ,j} \left (\frac{4}{3} D_x[v_x]_{i ,j}^{n} - \frac{2}{3} 
D_y[v_y]_{i ,j}^{n}\right ) \notag \\[2mm]
&-  (\alpha_{2}^{0}/\tau)_{i,j} A_t[s_{xx}]_{i,j}^{n},
\end{align}
\begin{align}\label{eq:s22}
&D_t[s_{yy}]_{i ,j}^{n} = (\mu)_{i ,j} \left (\frac{4}{3} D_y[v_y]_{i ,j}^{n} - \frac{2}{3} 
D_x[v_x]_{i ,j}^{n}\right ) \notag \\[2mm]
&- (\alpha_{2}^{0}/\tau)_{i,j}  A_t[s_{yy}]_{i,j}^{n},
\end{align}
\begin{equation}\label{eq:s12}
\begin{array}{c}
D_t[s_{xy}]_{i+1/2 ,j+1/2}^{n} = \left \{ \mu \right \}_{i+1/2 ,j+1/2} \left ( D_x[v_y]_{i+1/2 
,j+1/2}^{n} \right. \\ \left. + D_y[v_x]_{i+1/2 ,j+1/2}^{n}\right )
-  \left \{ \alpha_{2}^{0}/\tau  \right \}_{i+1/2 ,j+1/2} A_t[s_{xy}]_{i+1/2 ,j+1/2}^{n},
\end{array}
\end{equation}
\begin{align}\label{eq:p}
& D_t[P]_{i ,j}^{n}  = -  (K)_{i ,j}  \left (D_x[v_x]_{i ,j}^{n} +D_y[v_y]_{i ,j}^{n}\right ) - 
\notag\\
&-\left ( (\rho_{2}^{0}-\rho_{1}^{0})\alpha_{1}^{0}\alpha_{2}^{0}K/\rho^{0}\right )_{i ,j}\left (   
D_x[w_x]_{i ,j}^{n} +  D_y[w_y]_{i ,j}^{n} \right ),
\end{align}
\end{subequations}
where the effective medium parameters on the staggered grids are obtained as 
 volume arithmetic or harmonic averaging \cite{Moszo2002}:
\begin{align}\label{eq:notations}
&\left \langle f\right \rangle_{i+1/2,j}=(f_{i,j}+f_{i+1,j})/2,\notag \\
&\left \langle f\right \rangle_{i,j+1/2}=(f_{i,j}+f_{i,j+1})/2,\notag \\
&\left \{ f \right \}_{i+1/2 ,j+1/2}  = 4/\left(   
\dfrac{1}{f_{i,j}}+\dfrac{1}{f_{i+1,j}}+\dfrac{1}{f_{i,j+1}}+\dfrac{1}{f_{i+1,j+1}}\right).
\end{align}

The thus obtained 2D velocity-stress finite difference scheme is second-order accurate in 
both time and space.  Note that the approximation can be easily improved by using 
higher order operators.  The well-known Courant-Friedrichs-Lewy (CFL) stability 
criterion also holds for this case:  the time step must be 
chosen small enough in order that the fastest characteristic wave (P-wave) 
$C_{max}$ 
travel a distance smaller than the spatial discretization step:
$$
\Delta t C_{max}\sqrt{\frac{1}{{\Delta x}^{2}}+\frac{1}{{\Delta y}^{2}}}\leq 1.
$$

A forcing function $f(t,x,y)$ is introduced as the source term in the 
right-hand side of the pressure equation or the equations for the normal components 
of 
the deviatoric stress in system \eqref{stress.velocity}. In both cases, a 
volumetric-type source term is obtained. The source function is 
defined as the product of Dirac's delta function in space and Ricker's wavelet in time: 
\begin{equation}
f(t)=(1-2\pi ^{2} f_{0}^{2}(t-t_{0})^{2})exp[-\pi ^{2}f_{0}^{2}(t-t_{0})^{2}],
\end{equation}
where $f_{0}$ is the source peak frequency and $t_{0}$ is the wavelet delay.

No special care is taken to suppress the  outgoing waves with the help of absorbing boundary conditions (for example, PML). The simulation is stopped before the waves have reached the boundaries of the computational domain. 
The numerical experiments have been performed on a desktop computer with Intel(R) 
Core(TM) i7 3.60 GHz processor.

In the subsequent sections, we  will illustrate the main features of  
wavefield formation and propagation depending on porosity $\phi$, 
friction parameter $\theta_2$, source peak frequency $f_{0}$, and  shear relaxation time $\tau$. Before doing 
this, we start with considering the homogeneous dissipation-less case.

\subsection{Dependence on  porosity $\phi$.}

The computational domain $\Omega =[-0.65, 0.65]^{2}$  m was discretized with 
$ N_x \times N_y$ grid points, $N_x = N_y = 3250$, which amounts to 10 points 
per 
slow 
compressional wavelength in $\Omega$ for a source of central frequency 
$f_0=10^{5} $ Hz  and for various values of porosity $\phi$. The model 
parameters were taken from Table\,\ref{tab:parameters}. The source was located 
in the center of the computational 
domain. The propagation time was chosen to be equal to $T_{0}= 1.1\cdot10^{-4}$ s  with 
 source time delay $t_{0}=1/f_{0}$=$ 0.1\cdot10^{-4}$ s. 
The time step was chosen according to 
the classical Courant stability criterion for staggered grids with $CFL = 0.9$.

\begin{figure*}[!htbp]
\begin{subfigure}{0.5\linewidth}
\includegraphics[draft=false,width=0.7\textwidth]{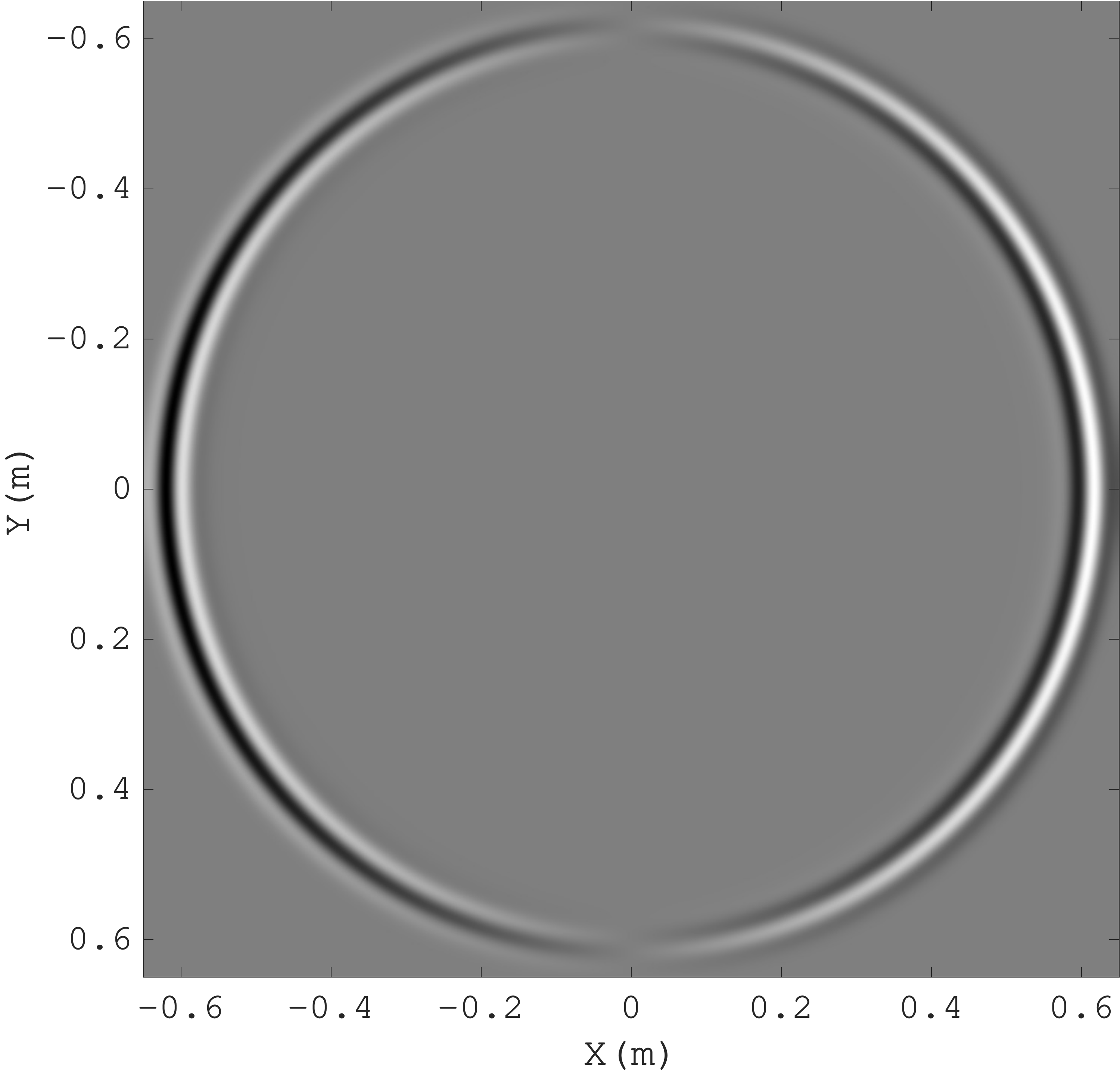}
\caption{}
\end{subfigure}
\hfill
\begin{subfigure}{0.5\linewidth}
\includegraphics[draft=false,width=0.8\textwidth]{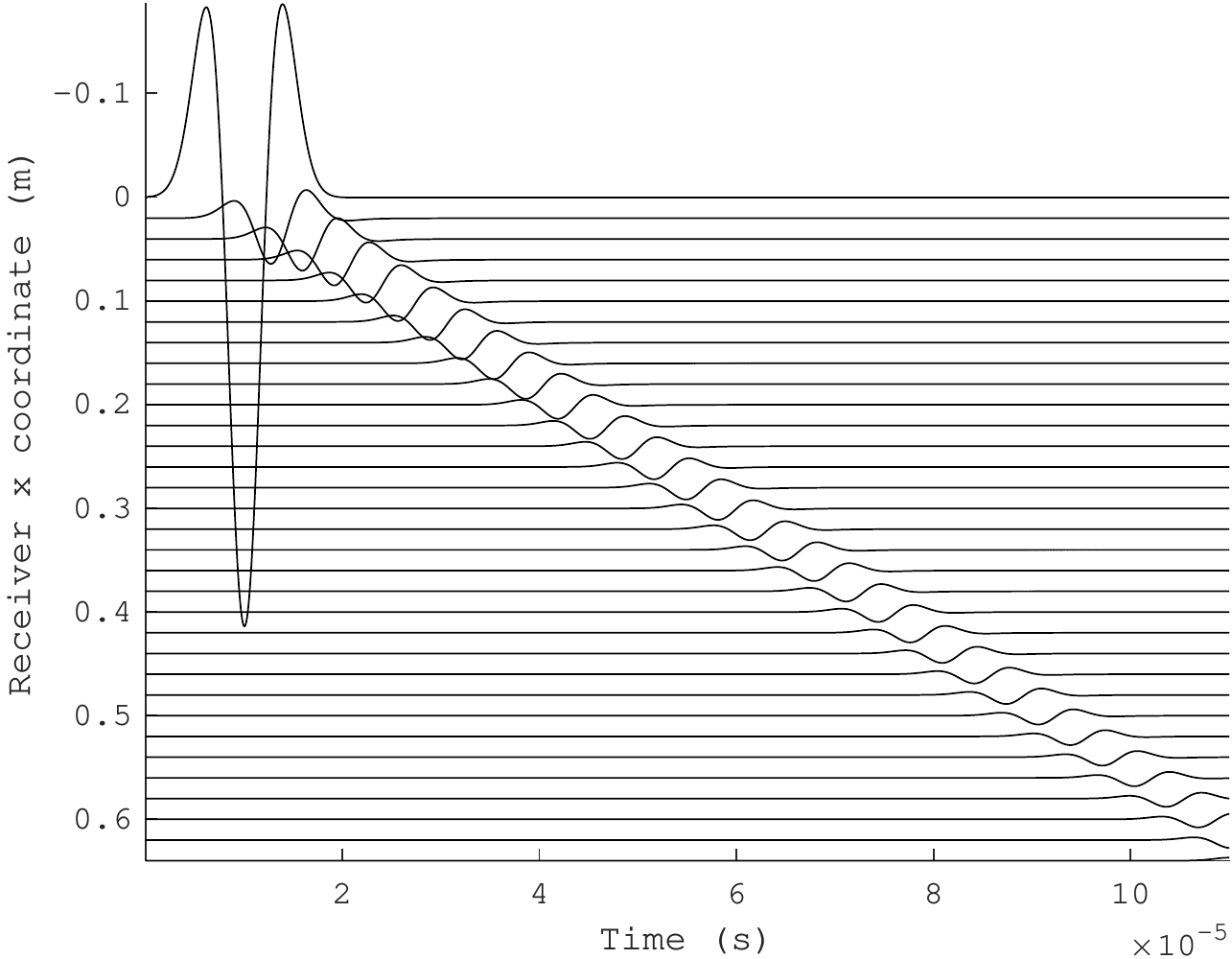}
\caption{}
\end{subfigure}%
\caption{
{
Wavefield in the pure elastic medium generated by the Ricker wavelet of a volumetric type. A snapshot at time $T_{0}= 1.1\cdot10^{-4}$ s (a) and seismogram (b) of the horizontal mixture 
velocity $v^1$  for source of central frequency $f_{0} =10^{5} $ Hz. }
}
\label{fig:porosity 0}
\end{figure*}

Fig.\,\ref{fig:porosity 0}(a) shows a snapshot of the mixture velocity $v^1$  
at time $T_{0}$, on the whole computational domain. The porosity parameter was 
chosen to be equal to $\phi$=0, which corresponds to the case of a pure elastic solid. 
In an elastic medium, only one fast P-wave with a velocity  equal to 6155 m/s 
is 
excited from a source of volumetric type. The seismogram in Fig.\,
\ref{fig:porosity 0}(b) confirms the occurrence of this wave with the predicted 
velocity. The receivers were located along the $ x $-axis starting from 
the source point towards the boundary with a uniform 
spacing between the receivers.

\begin{figure*}[!htbp]
\begin{subfigure}{0.5\linewidth}
\includegraphics[draft=false,width=0.7\textwidth]{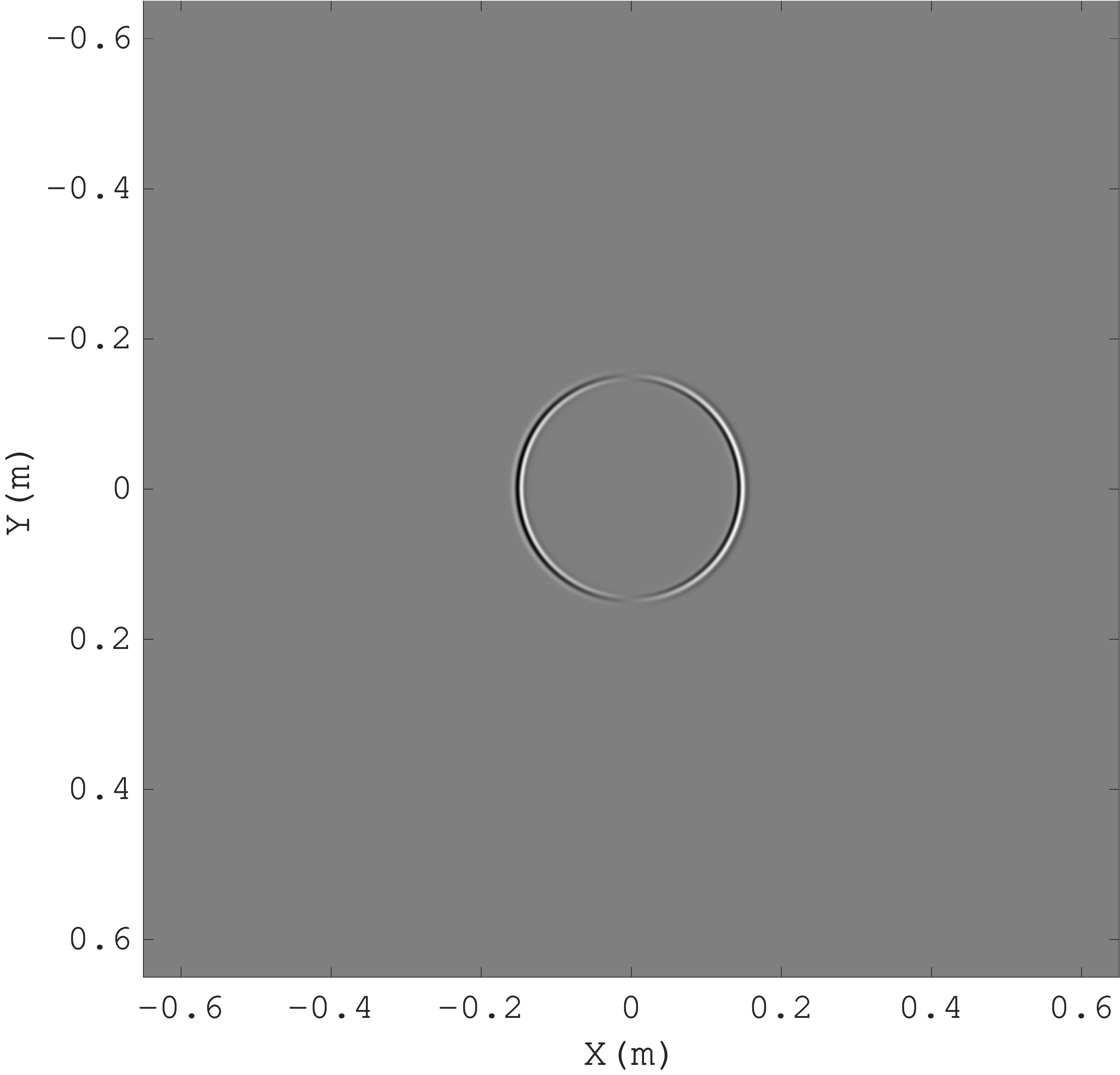}
\caption{}
\end{subfigure}
\hfill
\begin{subfigure}{0.5\linewidth}
\includegraphics[draft=false,width=0.8\textwidth]{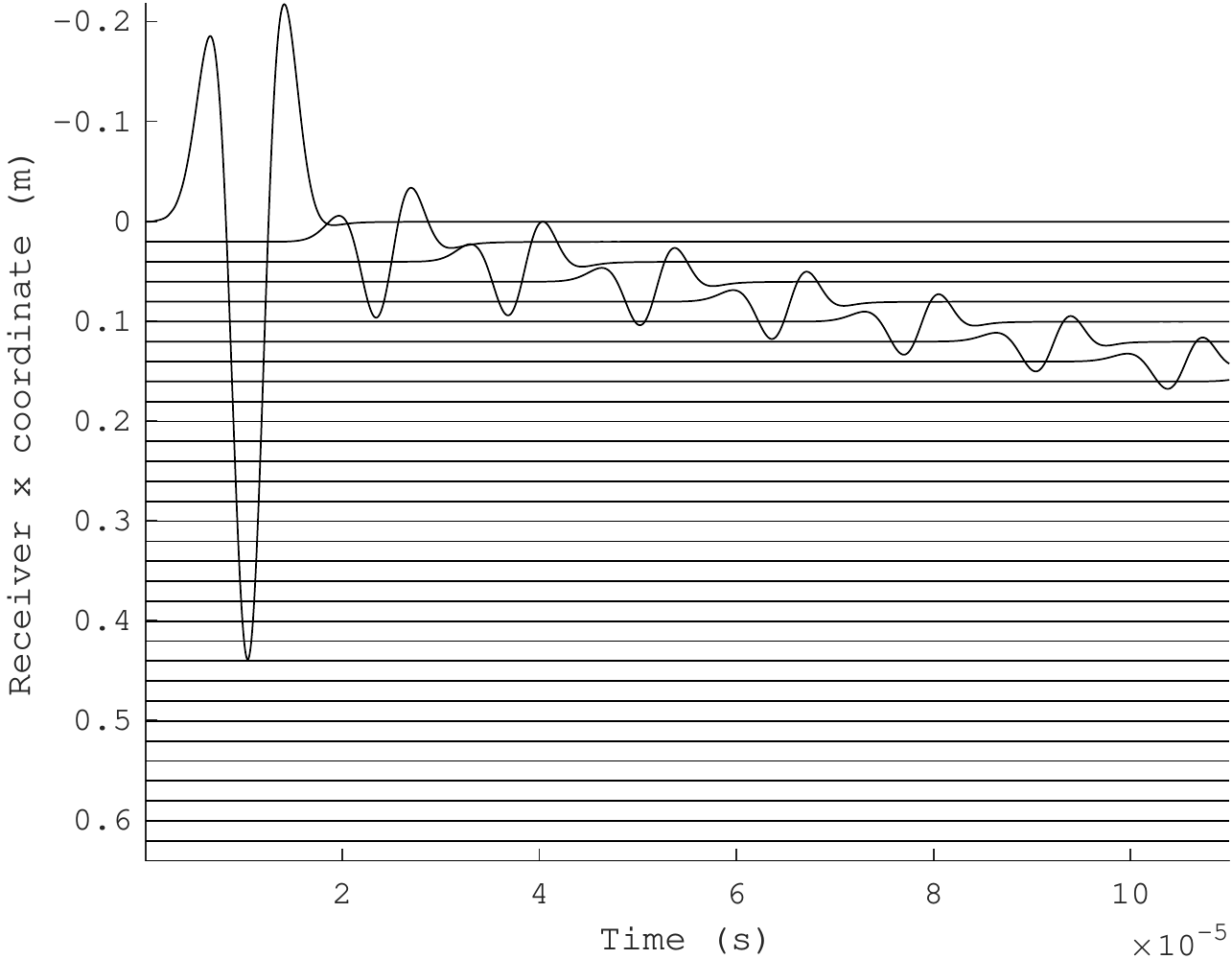}
\caption{}
\end{subfigure}%
\caption{
Wavefield in the pure fluid generated by the Ricker wavelet of a volumetric type.        A snapshot at time $T_{0}= 1.1\cdot10^{-4}$ s (a) and seismogram (b) of the mixture 
horizontal velocity $v^1$  for source of central frequency $f_{0} =10^{5} $ Hz.
}
\label{fig:porosity 1}
\end{figure*}

 Fig.\,\ref{fig:porosity 1} shows the results of calculations similar to the previous ones 
but with  porosity parameter $\phi$=1. This corresponds to the case of a pure 
liquid with one pressure wave with a velocity of 1500 m/s. The numerical 
propagation velocity can be easily estimated from the computed seismogram in  
Fig.\,\ref{fig:porosity 1}(b) to be exactly 1500 m/s.

\begin{figure*}
\begin{subfigure}{0.5\linewidth}
\includegraphics[draft=false,width=0.7\textwidth]{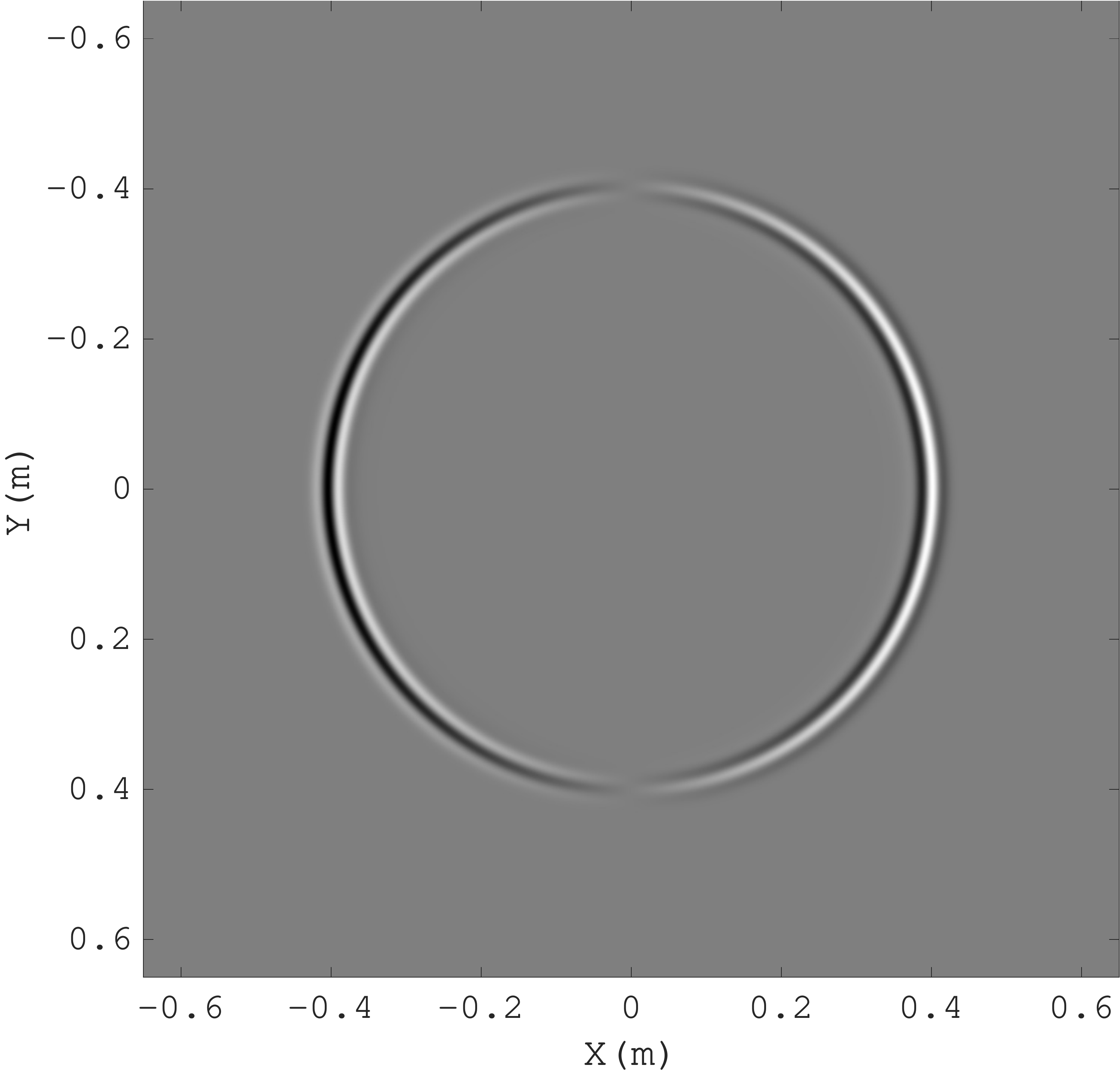}
\caption{}
\end{subfigure}
\hfill
\begin{subfigure}{0.5\linewidth}
\includegraphics[draft=false,width=0.8\textwidth]{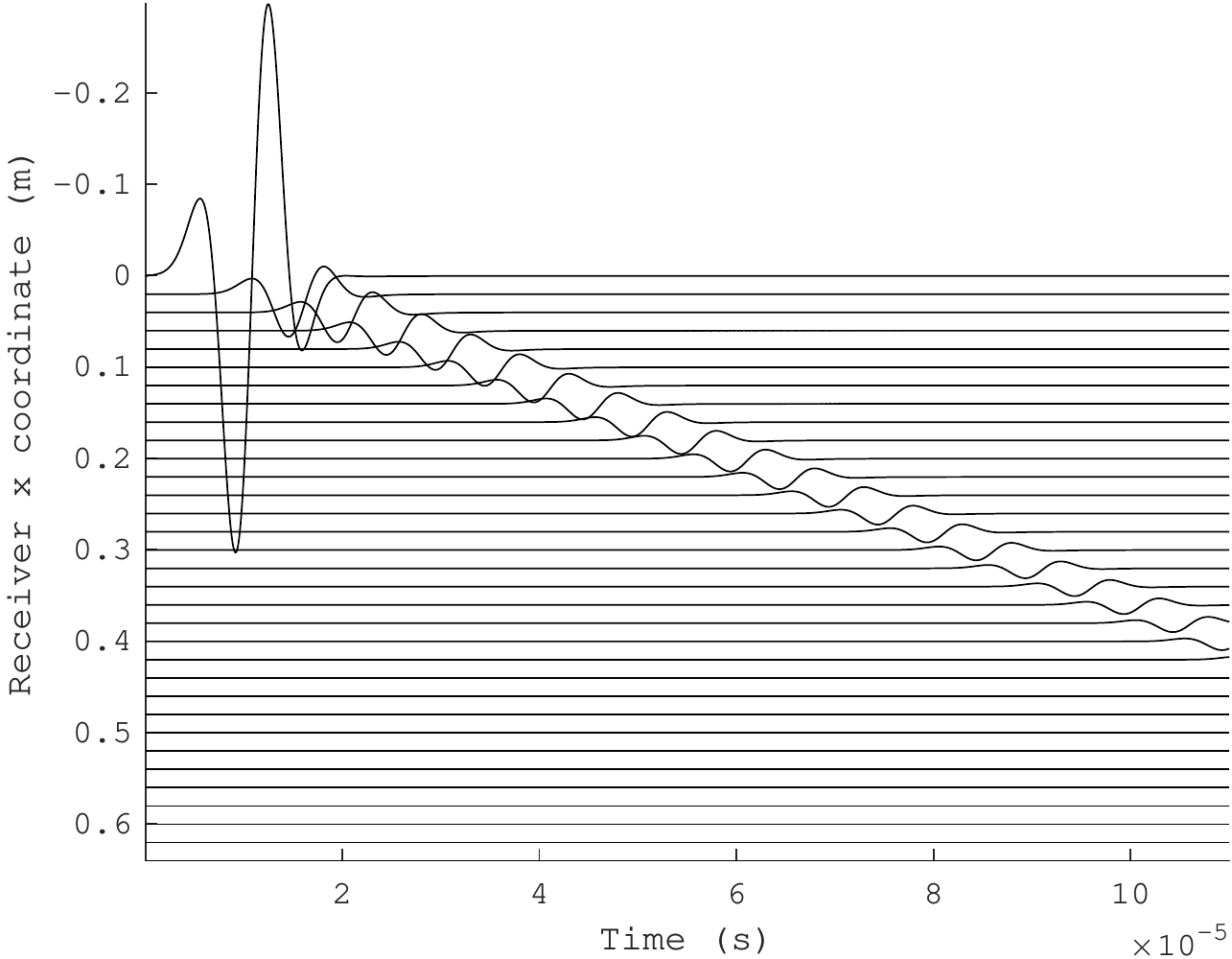}
\caption{}
\end{subfigure}%
\caption
{ Wavefield in the porous medium with porosity $\phi=0.5$ generated by the Ricker wavelet of a 
volumetric type. A snapshot at time $T_{0}= 1.1\cdot10^{-4}$ s (a) and seismogram (b) of the 
mixture horizontal velocity $v^1$  for source of central frequency $f_{0} =10^{5}$ Hz.
}
\label{fig:porosity 05}
\end{figure*}

\begin{figure*}
\begin{subfigure}{0.5\linewidth}
\includegraphics[draft=false,width=1.2\textwidth]{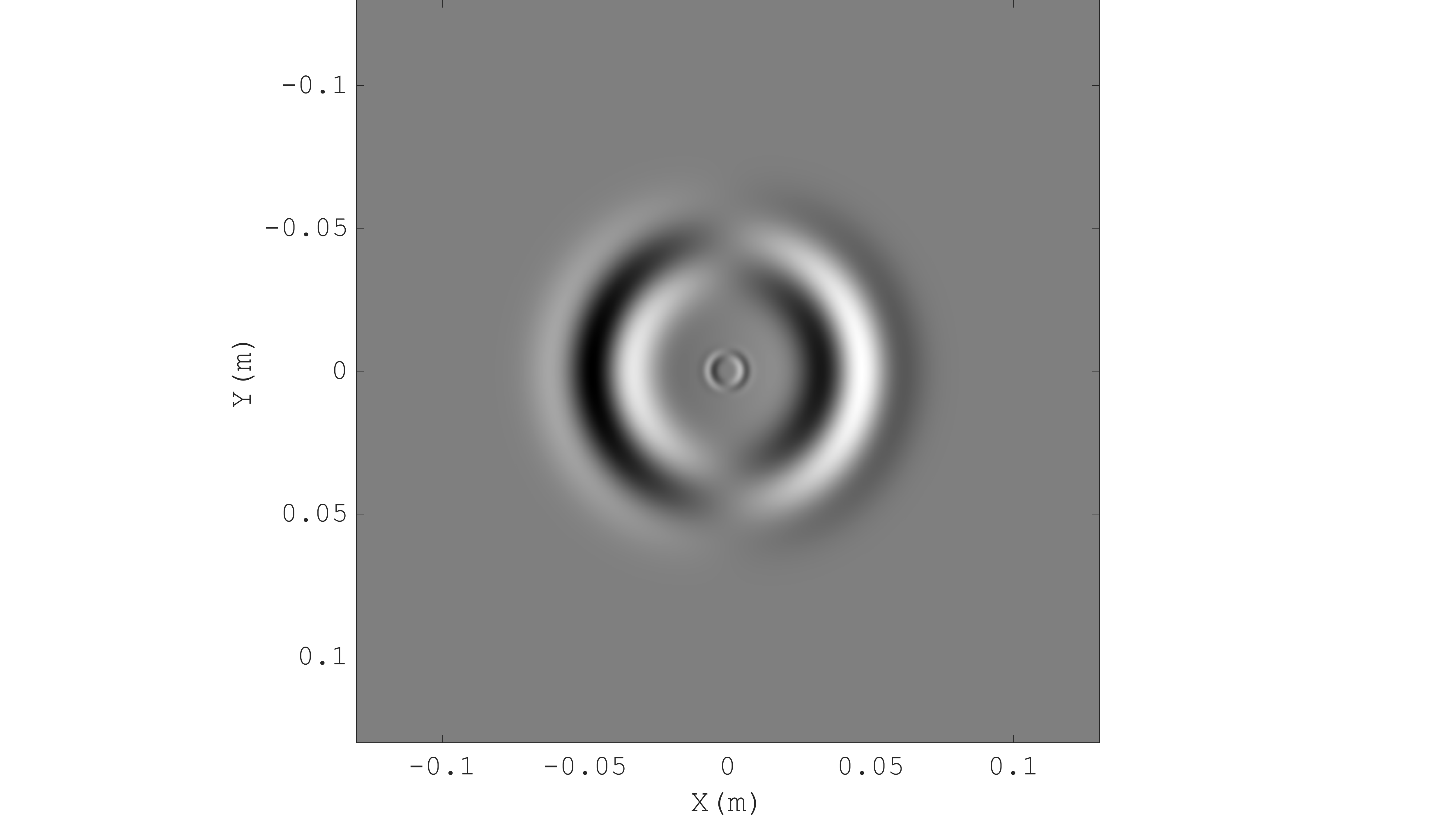}
\caption{}
\end{subfigure}
\hfill
\begin{subfigure}{0.5\linewidth}
\includegraphics[draft=false,width=0.8\textwidth]{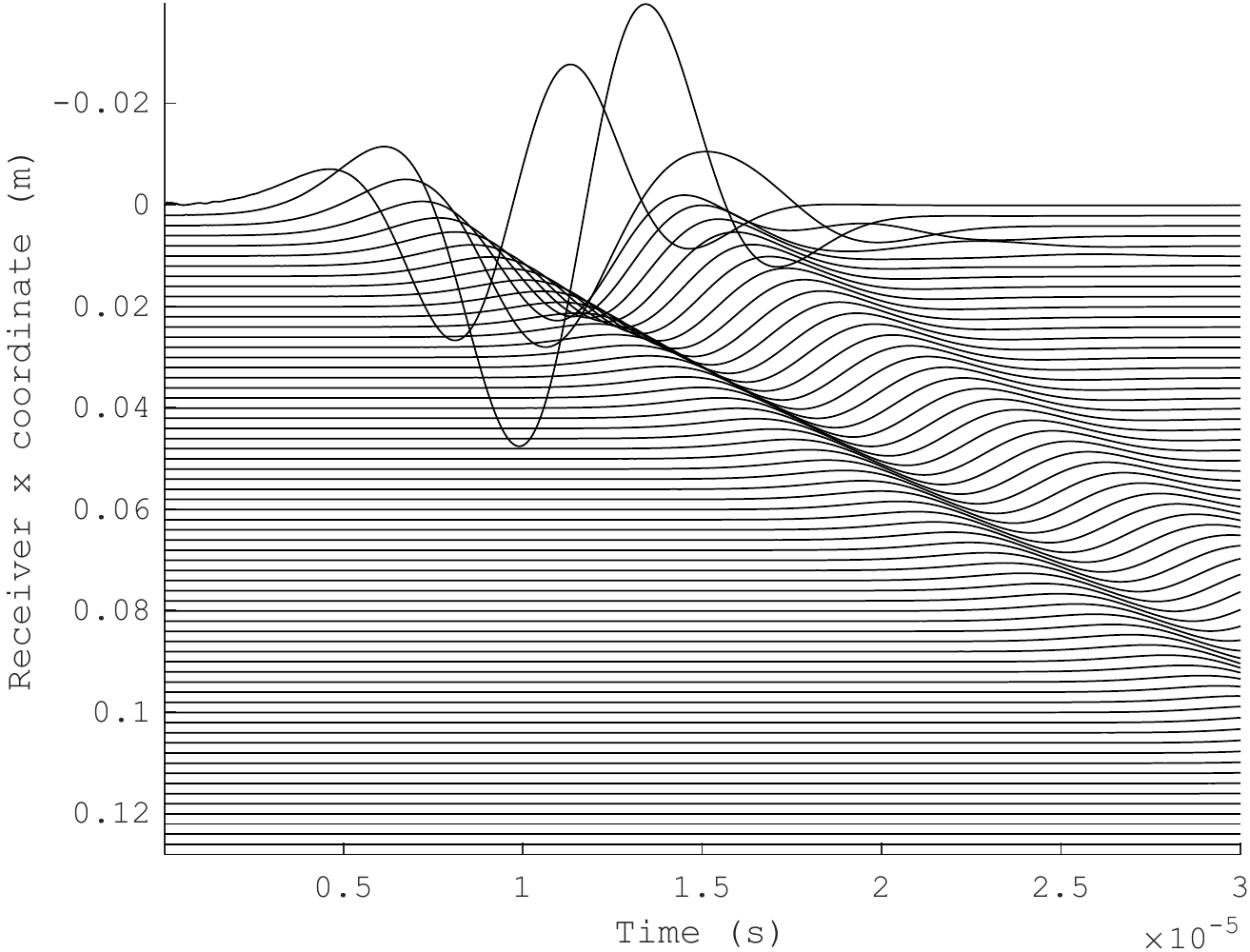}
\caption{}
\end{subfigure}%
\caption{
{Wavefield in the porous medium with porosity $\phi=0.5$ generated by the Ricker wavelet of a 
volumetric type. A zoom of snapshot at time $1\cdot10^{-5}$ s (a) and seismogram (b) of the mixture 
horizontal
velocity $v^1$  for source of central frequency $f_{0} =10^{5} $ Hz.}
}
\label{fig:porosity 05 zoom}
\end{figure*}

\begin{figure*}[!htbp]
	\begin{center}
		\includegraphics[draft=false,width=1.0\textwidth]{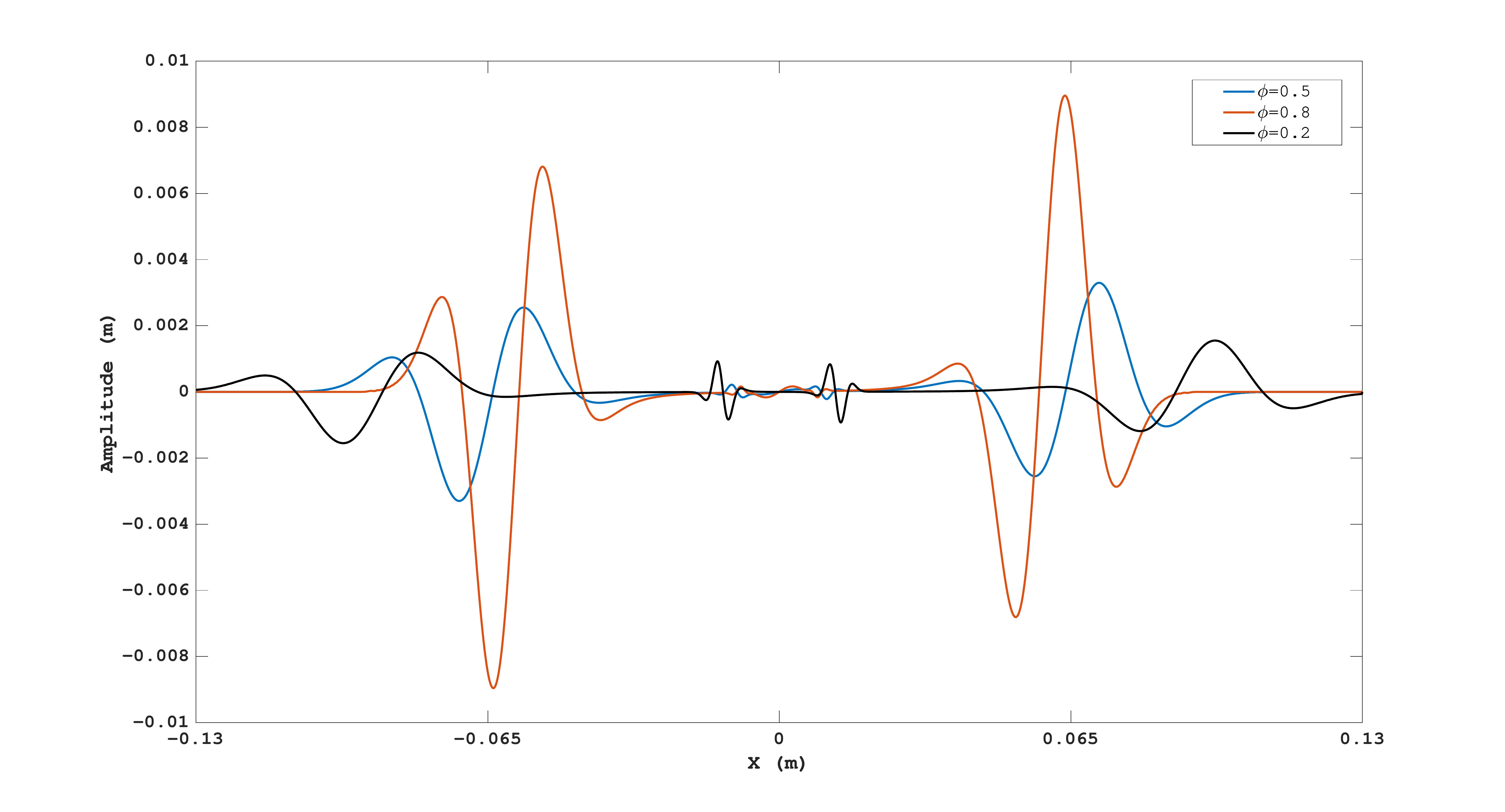}
	\end{center}
\caption{
The distribution of the mixture horizontal velocity $v^1$ generated by the Ricker wavelet of a 
volumetric type along the $x$-axis at  
time $2.1\cdot10^{-5}$ s and $y=0$  for several values of 
porosity $\phi$: $\phi=0.2$ (black), $\phi=0.5$ (blue), $\phi=0.8$ 
(red).
}
\label{fig:porosity different}
\end{figure*}

The simulation results for  porosity parameter $\phi$=0.5 are shown in 
Fig.\,\ref{fig:porosity 05}. The fast P-wave velocity is estimated to be 4100 m/s 
from the
computed seismogram in  Fig.\,\ref{fig:porosity 05}(b), which is consistent with 
the data from Table\,\ref{tab:parameters}. In the snapshot of Fig.\,\ref{fig:porosity 
05}(a) we do not observe the predicted slow P-wave because it is  completely attenuated, 
not visible at this time, and its amplitude is 
very 
small. However, if we 
zoom in the image, we will be able to see this slow P-wave in 
Fig.\,\ref{fig:porosity 05 zoom}.
 The amplitude variation  for several values of $\phi$ can also be 
 seen in Fig.\,\ref{fig:porosity different}, where the wavefield distribution along the 
 $x$-axis for $y=0$ is presented. 

Summarizing  the results of numerical experiments of this section, we conclude that 
by varying porosity $\phi$ in system \eqref{stress.velocity}, it is 
possible to correctly describe the three states of the medium: liquid, solid, and 
poroelastic.

\subsection{Dependence on friction $ \theta_2 $.}

In this section, we study the behavior and properties of the fast and slow 
P-waves, depending on the parameter $\theta_2$. This parameter is present as 
the denominator in the right-hand side of the second equation in system 
\eqref{stress.velocity}. By analogy with Biot's model, $\theta_2$ can be 
viewed as a friction parameter, because it controls interfacial friction 
in a multiphase medium and leads to  wave dispersion and attenuation.

\begin{figure*}[!htbp]
\begin{subfigure}{0.3\linewidth}
\centering
\includegraphics[draft=false,width=1\textwidth]{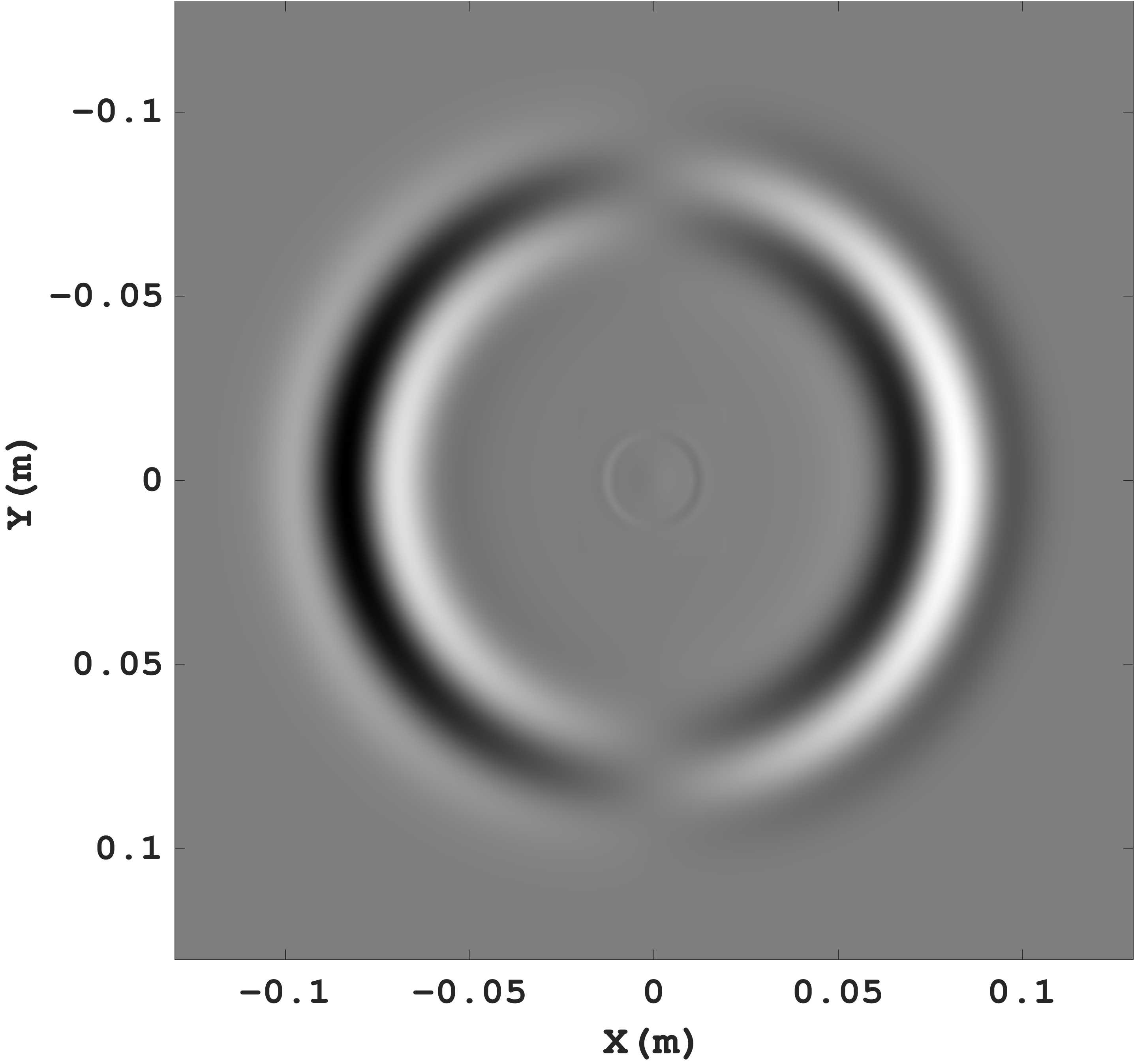}
\caption{$ \quad\quad \theta_2=3.36\cdot10^{-7} $}
\end{subfigure}
\hfill
\begin{subfigure}{0.3\linewidth}
\includegraphics[draft=false,width=1\textwidth]{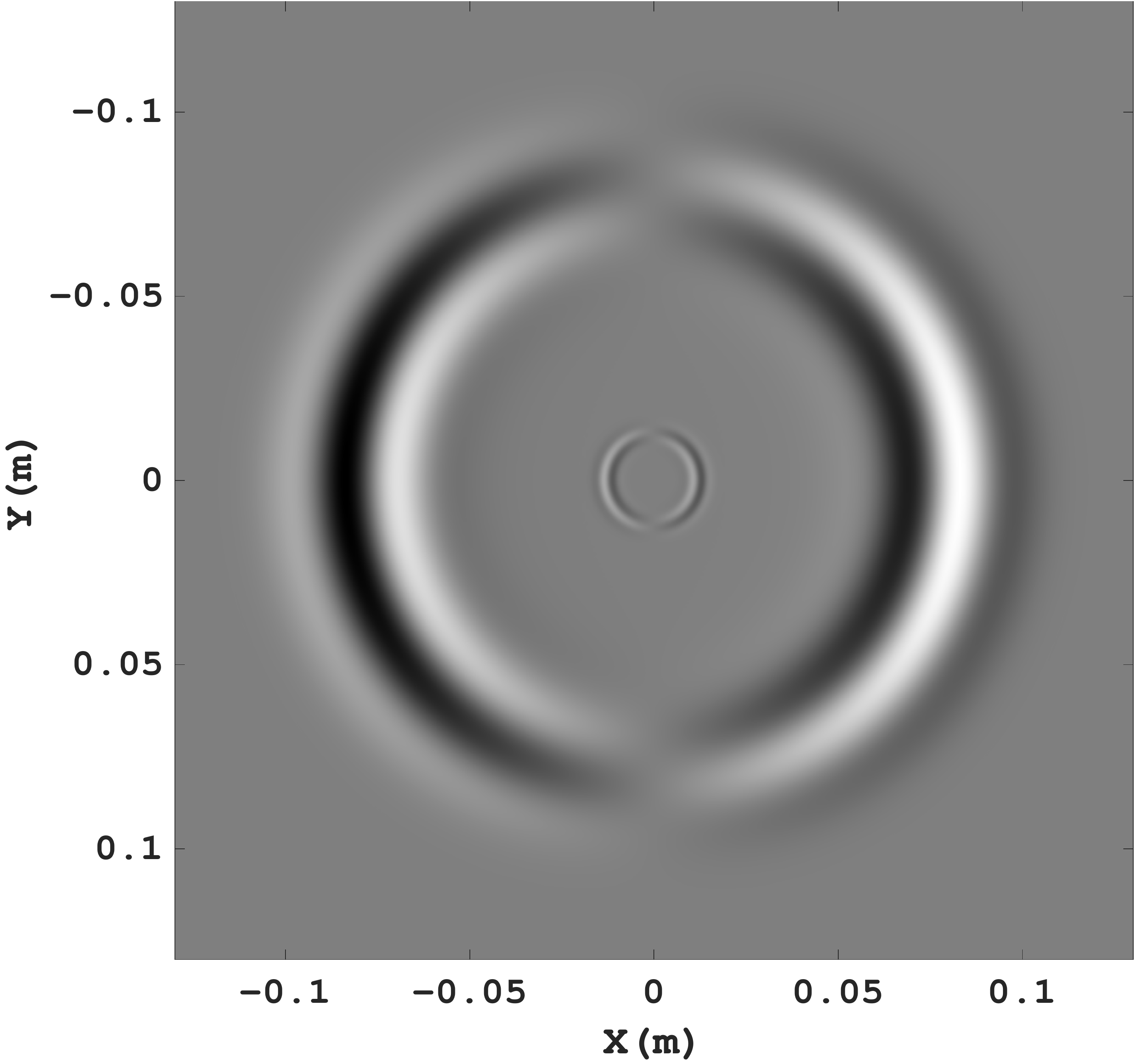}
\caption{$ \quad\quad \theta_2=2\cdot3.36\cdot10^{-7} $}
\end{subfigure}%
\hfill
\begin{subfigure}{0.3\linewidth}
\includegraphics[draft=false,width=1\textwidth]{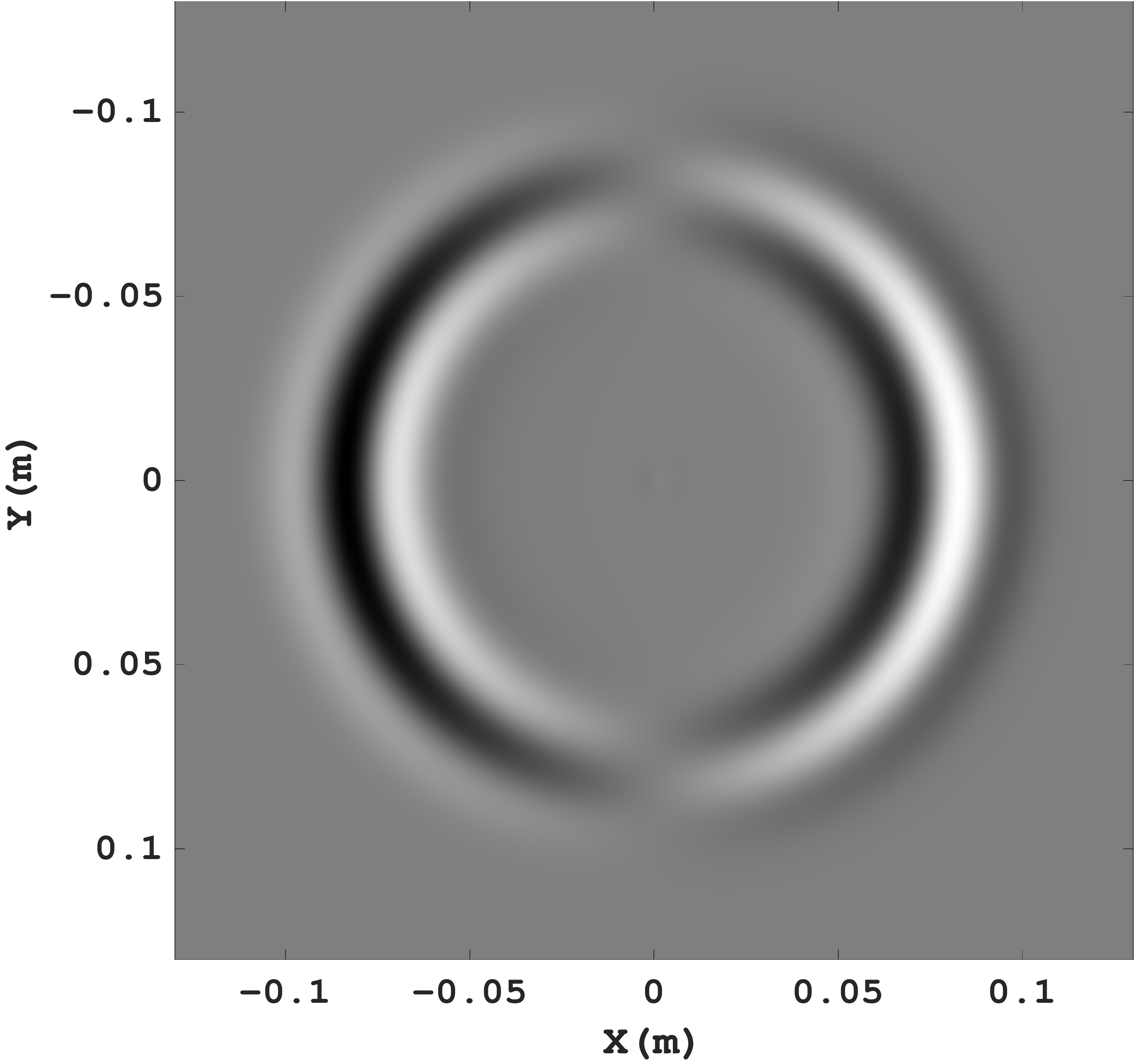}
\caption{$ \quad\quad \theta_2=1/2\cdot3.36\cdot10^{-7} $}
\end{subfigure}%
\caption{
Wavefield in the porous medium with porosity $\phi=0.5$ generated by the Ricker wavelet of a 
volumetric type.
Snapshots at time $2.1\cdot10^{-5}$ s of the horizontal mixture velocity $v^1$ for several values 
of $ \theta_2 $.
}
\label{fig:compare_friction}
\end{figure*}

\begin{figure*}[!htbp]
	\begin{center}
		\includegraphics[draft=false,width=1.0\textwidth]{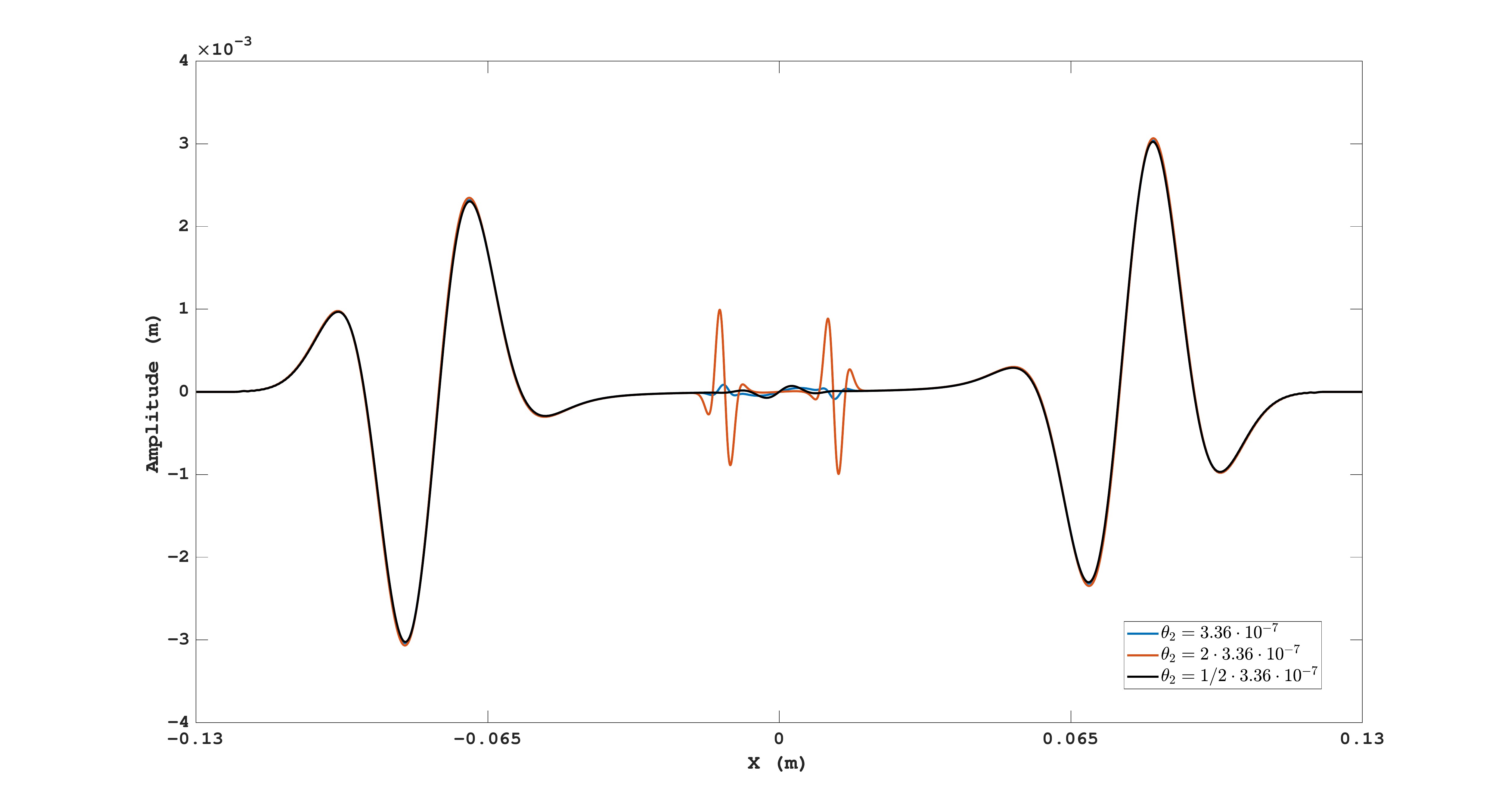}
	\end{center}
	\caption{A comparison at time $2.1\cdot10^{-5}$ s of values of the mixture velocity 
	$v^1$ for several values of $ \theta_2 $: $ \theta_2 
	=3.36\cdot10^{-7}$ (blue), $ \theta_2=2 \cdot 3.36\cdot10^{-7} $ (red), $ 
	\theta_2=1/2 \cdot 3.36\cdot10^{-7} $ (black).}
	\label{fig:compare_friction_line}
\end{figure*}

Consider the same homogeneous numerical model as in the previous sections with 
$ \theta_2$ equal to $3.36\cdot10^{-7}$ from Table\,\ref{tab:parameters}. Let 
us observe the behavior of the  P-waves if we increase or decrease 
this parameter twice. Fig.\,\ref{fig:compare_friction}(a-c) shows snapshots of 
the mixture velocity $v^1$ for these three values of $ \theta_2$. 
Significant wave amplitude variations are observed only for slow P-wave. More 
detailed variations of the amplitudes can be seen in 
Fig.\,\ref{fig:compare_friction_line}, where the wavefield along the 
$x$-axis 
at $y=0$ is presented. This figure demonstrates an increase in 
the 
amplitude of  the slow P-wave with increasing $\theta_2$, 
and vice versa. A change in the form of slow waves is also observed, while 
the fast wave remains almost unchanged.

Summarizing  the results of numerical experiments of this section, we conclude 
that by varying the parameter $\theta_2$ in system \eqref{stress.velocity} it 
is possible to affect the amplitude and  propagation velocity of the slow 
P-wave. One of the interesting applications, in our opinion, can be the 
solution of the inverse problem of determining the coefficient $\theta_2$ by 
analyzing the ratio of the amplitudes of the fast and slow waves in a field 
experiment.

\subsection{Dependence on frequency $f_{0}$.}\label{sec.frequency}

\begin{figure*}[!htbp]
\begin{subfigure}{0.3\linewidth}
\includegraphics[draft=false,width=1\textwidth]{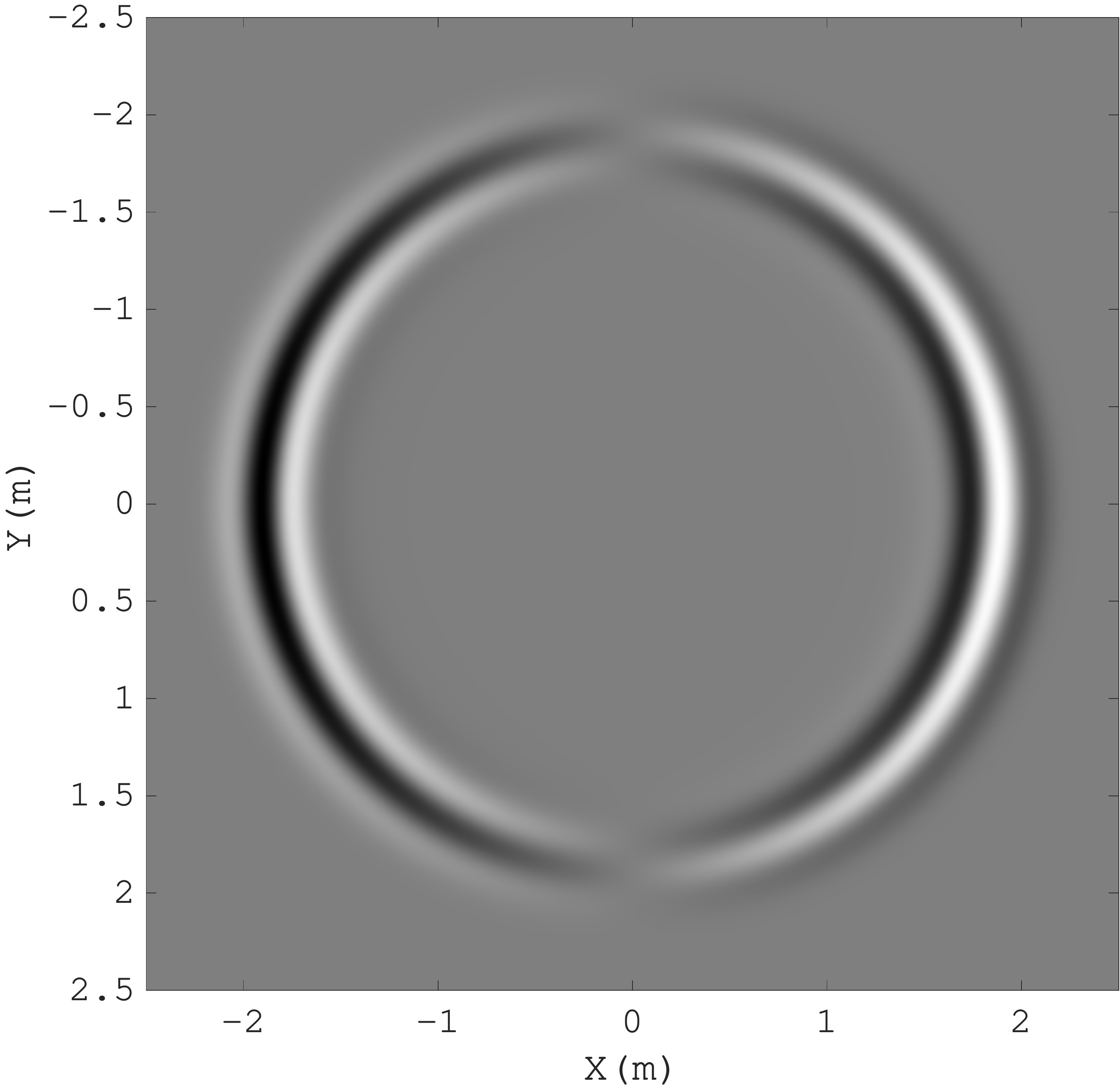}
\caption{$\quad\quad f_0=10^{4} $}
\end{subfigure}
\hfill
\begin{subfigure}{0.3\linewidth}
\includegraphics[draft=false,width=1\textwidth]{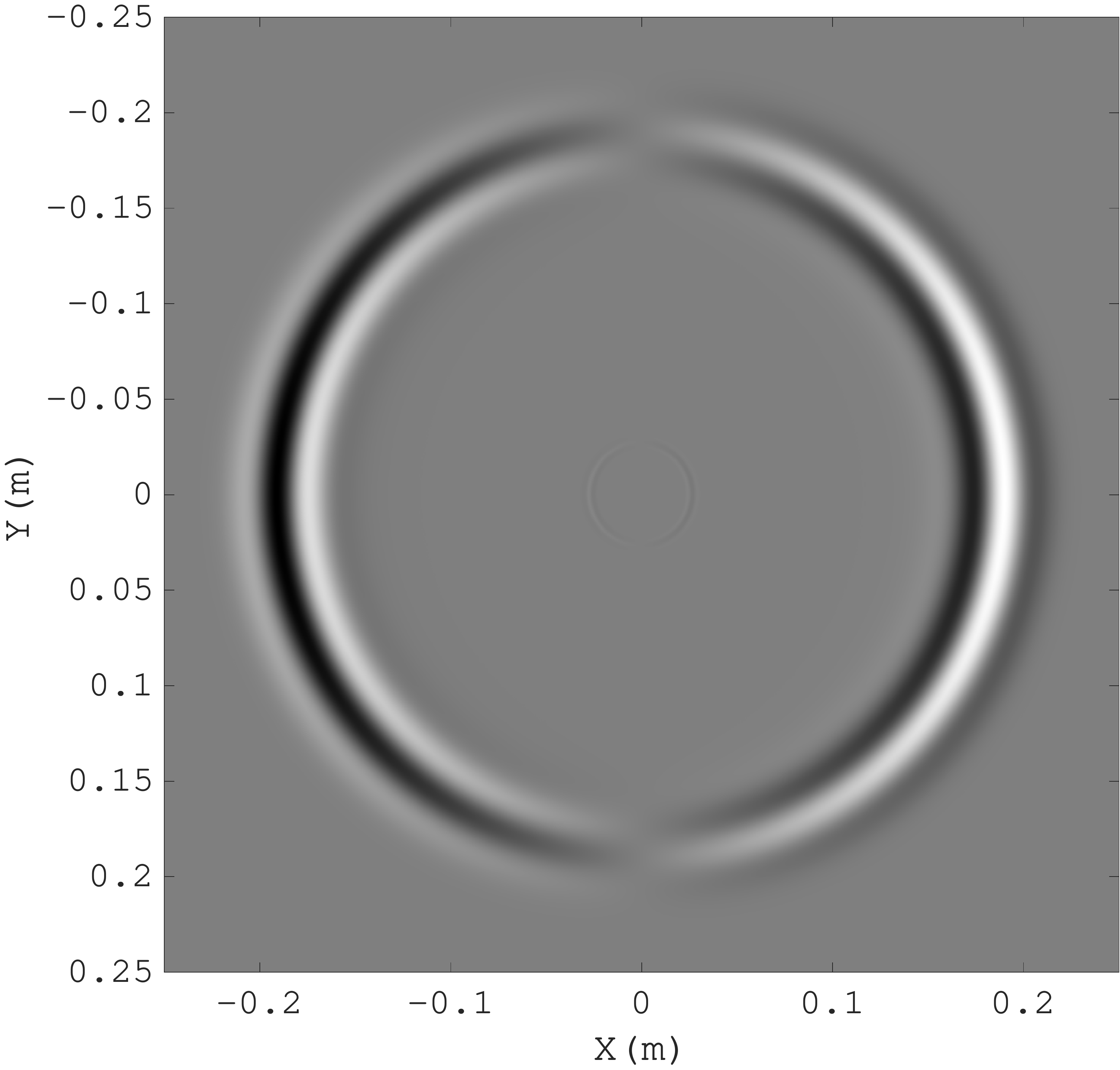}
\caption{$\quad\quad f_0=10^{5} $}
\end{subfigure}%
\hfill
\begin{subfigure}{0.3\linewidth}
\includegraphics[draft=false,width=1\textwidth]{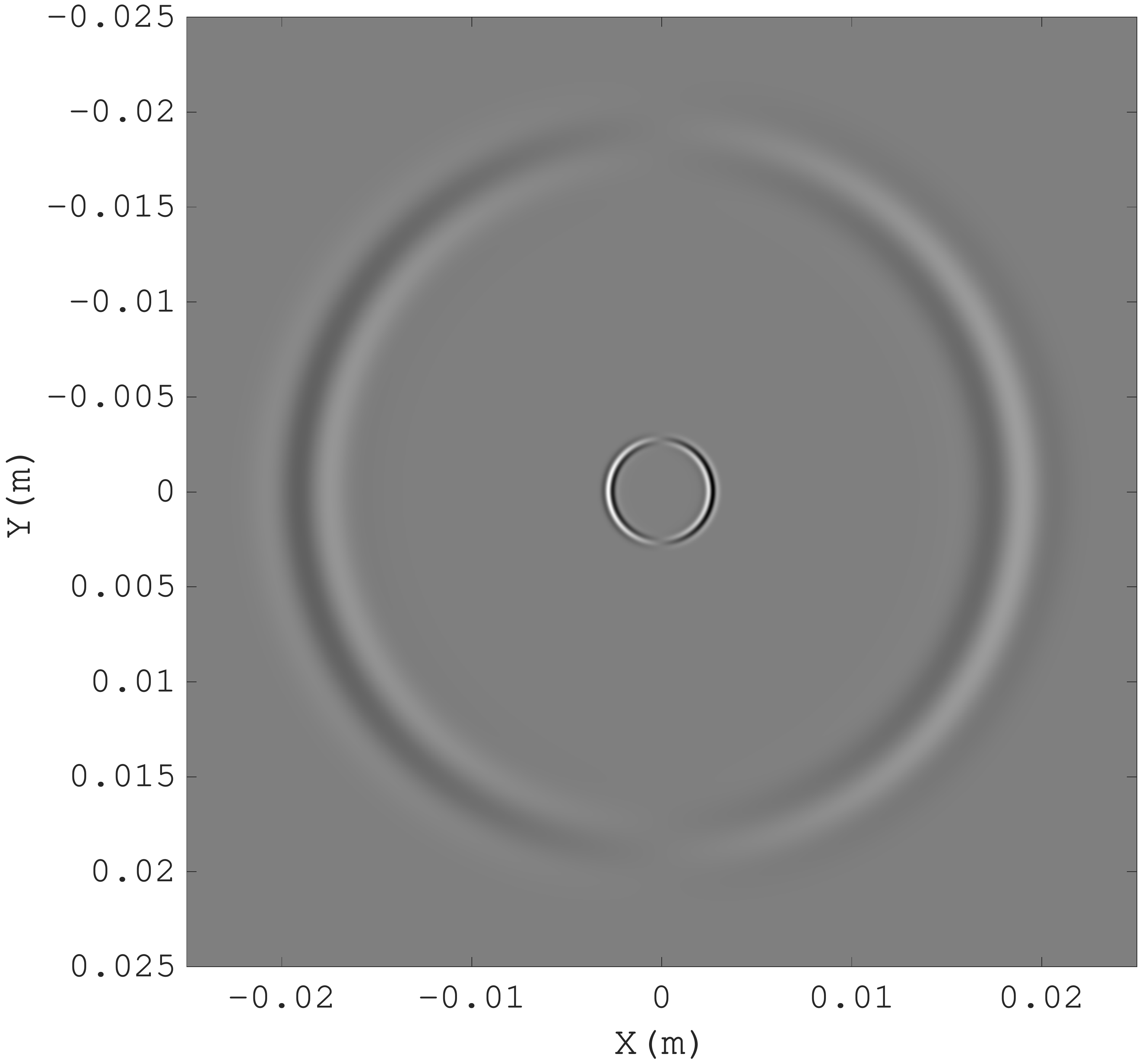}
\caption{$\quad\quad f_0=10^{6} $}
\end{subfigure}%
\caption{Wavefield in the porous medium with porosity $\phi=0.5$ generated by the Ricker wavelet of 
a volumetric type.
Snapshots of the horizontal mixture velocity $v^1$ for several frequencies: 
$f_0=10^{4}$ Hz (a), $f_0=10^{5}$ Hz (b), $f_0=10^{6}$ Hz (c). 
}
\label{fig:compare_frewuency}
\end{figure*}

\begin{figure*}[!htbp]
\begin{subfigure}{0.3\linewidth}
\includegraphics[draft=false,width=1\textwidth, 
height=0.15\textheight]{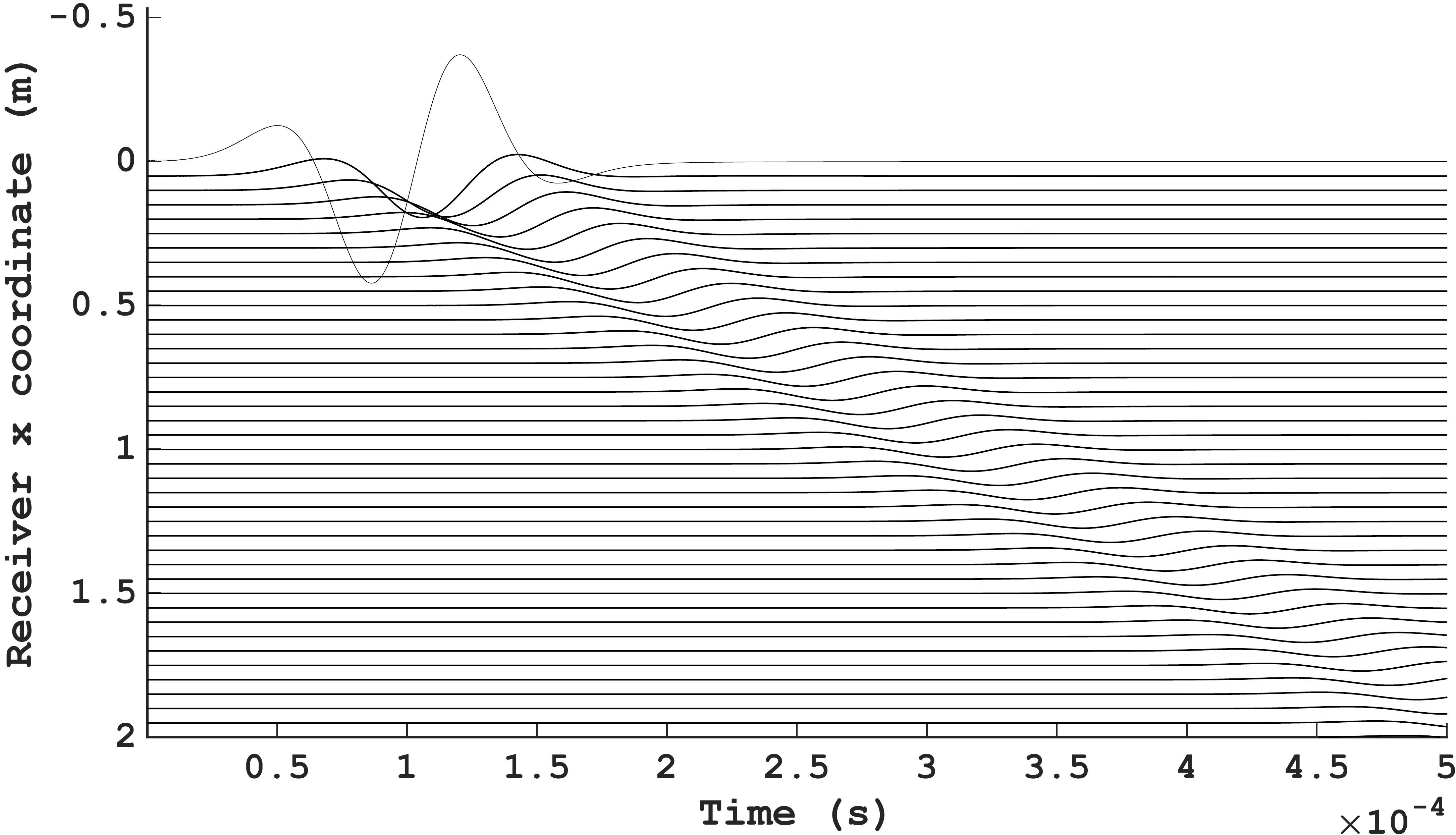}
\caption{$\quad\quad f_0=10^{4} $}
\end{subfigure}
\hfill
\begin{subfigure}{0.3\linewidth}
\includegraphics[draft=false,width=1\textwidth, 
height=0.15\textheight]{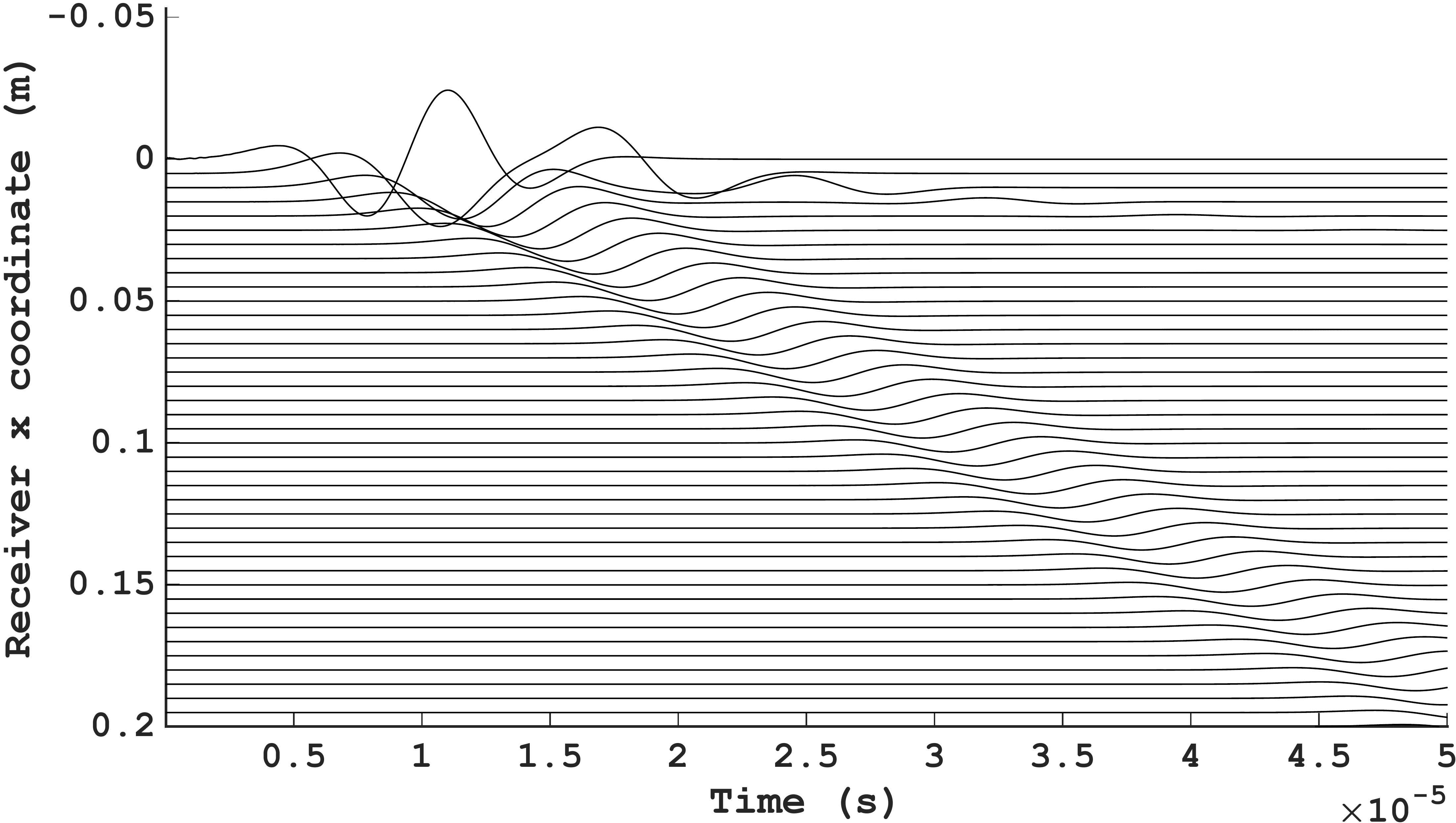}
\caption{$\quad\quad f_0=10^{5} $}
\end{subfigure}%
\hfill
\begin{subfigure}{0.3\linewidth}
\includegraphics[draft=false,width=1\textwidth, 
height=0.15\textheight]{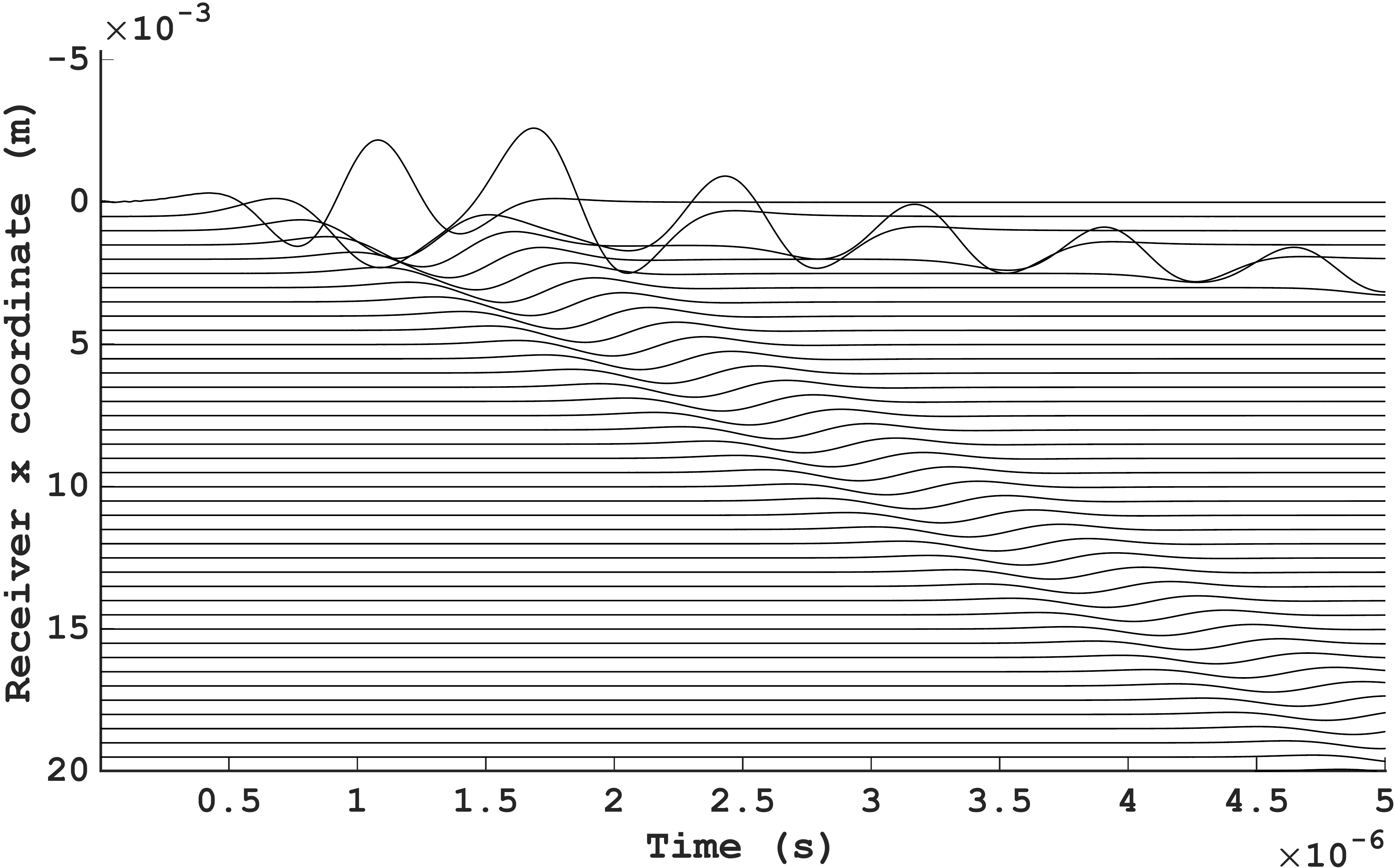}
\caption{$\quad\quad f_0=10^{6} $}
\end{subfigure}%
\caption{ 
Wavefield in the porous medium with porosity $\phi=0.5$ generated by the Ricker wavelet of a 
volumetric type.
Seismograms of the horizontal mixture velocity $v^1$ for several frequencies: $f_0=10^{4}$ Hz (a), 
$f_0=10^{5}$ Hz (b), $f_0=10^{6}$ Hz(c). 
}
\label{fig:compare_frewuency_trace}
\end{figure*}

In this section, we perform a study of dispersion of the wave velocity depending on the 
source 
peak frequency $f_{0}$  based on the same homogeneous numerical model 
with 
parameters from Table \ref{tab:parameters} as in the previous sections. 
The dispersion curves in Fig.\,\ref{fig:diff.grain.modulus} show that the main 
velocity changes are in the range of $10^{4}-10^{6}$ Hz. Because of a difference of 
three orders of magnitude in the frequency range, our comparison will be made 
not 
for a single computational domain, but for three different 
domains. More specifically, we use a ten orders of magnitude scaling of  both space and 
time. Fig.\,\ref{fig:compare_frewuency} presents 
snapshots, and Fig.\,\ref{fig:compare_frewuency_trace} presents seismograms of 
the mixture velocity $v^1$ for several frequencies. For each time frequency, 
a snapshot is recorded at time $5/{f_{0}}$ (including a shift wavelet 
delay of $1/{f_{0}}$) for square domains with a side of $ 5 $ m (for 
$f_{0}=10^{4}$), $0.5$  m (for 
$f_{0}=10^{5}$), and $0.05$  m (for $f_{0}=10^{6}$). The time 
and size are chosen in such a way that in the isotropic elastic case 
we can obtain three identical snapshots. As expected in the poroelastic 
case, 
we 
observe wavefield differences which are most clearly seen in the seismograms: the 
lower the frequency, the stronger the dispersion and attenuation of the slow 
P-wave. 
To estimate the phase velocity, we use 
a spectral ratio technique 
\cite{Gurevich2015}, \cite{Caspari2019}
and obtain a 
phase velocity $v_{p}=420 $ m/s for a frequency $f_{0}=10^{5}$ Hz and
$v_{p}=670 $ m/s  for the frequency $f_{0}=10^{6}$ Hz. 
It is not possible to make 
similar estimation for a frequency $f_{0}=10^{4}$ due to the fact that the slow wave has a low amplitude. We 
can also conclude that the attenuation of the slow P-wave is sufficiently strong even 
for high frequencies.

\subsection{Dependence on shear relaxation time.}

The aim of this section is to show that there is an additional mechanism of 
energy dissipation embedded in system \eqref{stress.velocity} and controlled 
by the relaxation parameter $\tau$ in the right-hand side of equation
\eqref{sij}. A proper choice of the relaxation time $ \tau $ allows one to 
model 
irreversible (elastoplastic) deformations in the solid matrix, e.g. 
\cite{HYP2016,Hyper-Hypo2019}, or viscous flows \cite{DPRZ2016}.

All the previous numerical examples were simulated without relaxation of tangential 
shear stresses, that is, the right-hand side in \eqref{sij} vanished. Formally, 
this corresponds to the case of $\tau=\infty$. It is quite obvious that, for 
finite values of
$ \tau $, the mechanism of relaxation of tangential stresses  provides 
an additional ability of the model to 
control wave attenuation. To this end, we again 
consider the
example from the previous Section\,\ref{sec.frequency} for a frequency 
$f_0=10^{4}$ Hz and compare 
the solution with a similar test, but with allowance for  relaxation of 
tangential stresses.
\begin{figure*}[t]
	\begin{center}
\includegraphics[draft=false,width=0.7\textwidth]{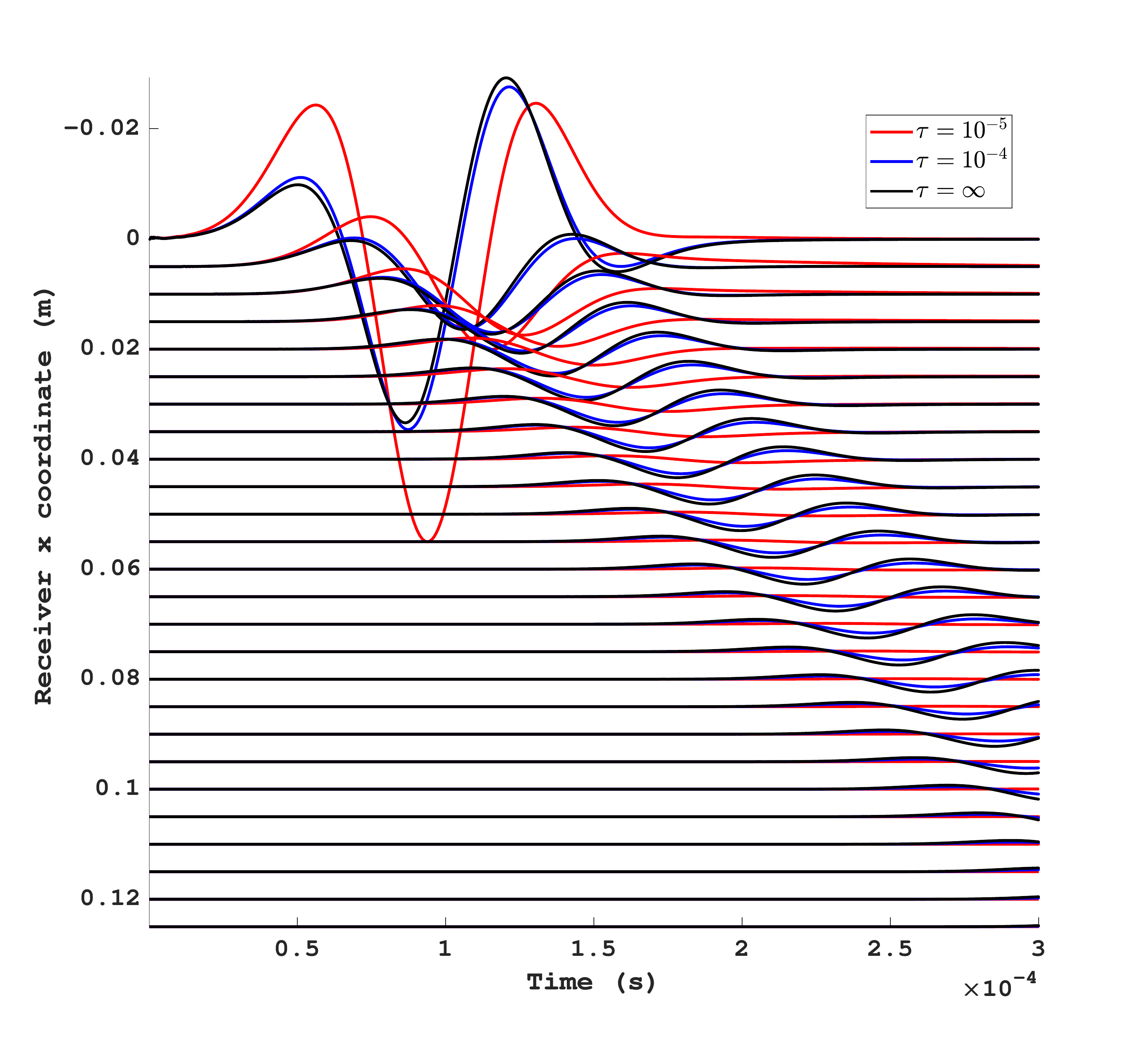}
	\end{center}
	\caption{{
	Comparison of seismograms of the horizontal mixture velocity $v^1$  for several 
	values of shear relaxation time $\tau$:
	no relaxation (black), $\tau=10^{-5}$ (red), $\tau=10^{-4}$ (blue).}
	}
	\label{fig: Compare_dissip}
\end{figure*}
As expected, a comparison of the seismograms in Fig.\,\ref{fig: Compare_dissip} shows that relaxation of tangential stresses leads to dispersion and attenuation of seismic waves. On the distant receivers, it is seen that the smaller $\tau$, the greater the attenuation. On the nearer receivers, dispersion probably has a predominant effect.

\subsection{Layered media.}

This test illustrates the effects of the interfaces  between pure fluid, poroelastic 
and 
pure elastic media on an example of a three-layered medium with the media 
parameters from Table 
\ref{tab:parameters}. The size of the computational domain is $0.025$\,m in the 
$ x $ and $ y $ directions. The upper layer is water, the lower layer is an
elastic 
medium and in the middle, from $-0.005$ m to $0.005$ m, there is a poroelastic 
layer with porosity $\phi$=0.2. A source of central frequency $f_0=10^{6}$ Hz 
is located in the water layer at the point $x=0, y=-0.01$. 

\begin{figure*}[t]
	\begin{center}
	\includegraphics[draft=false,width=0.8\textwidth]{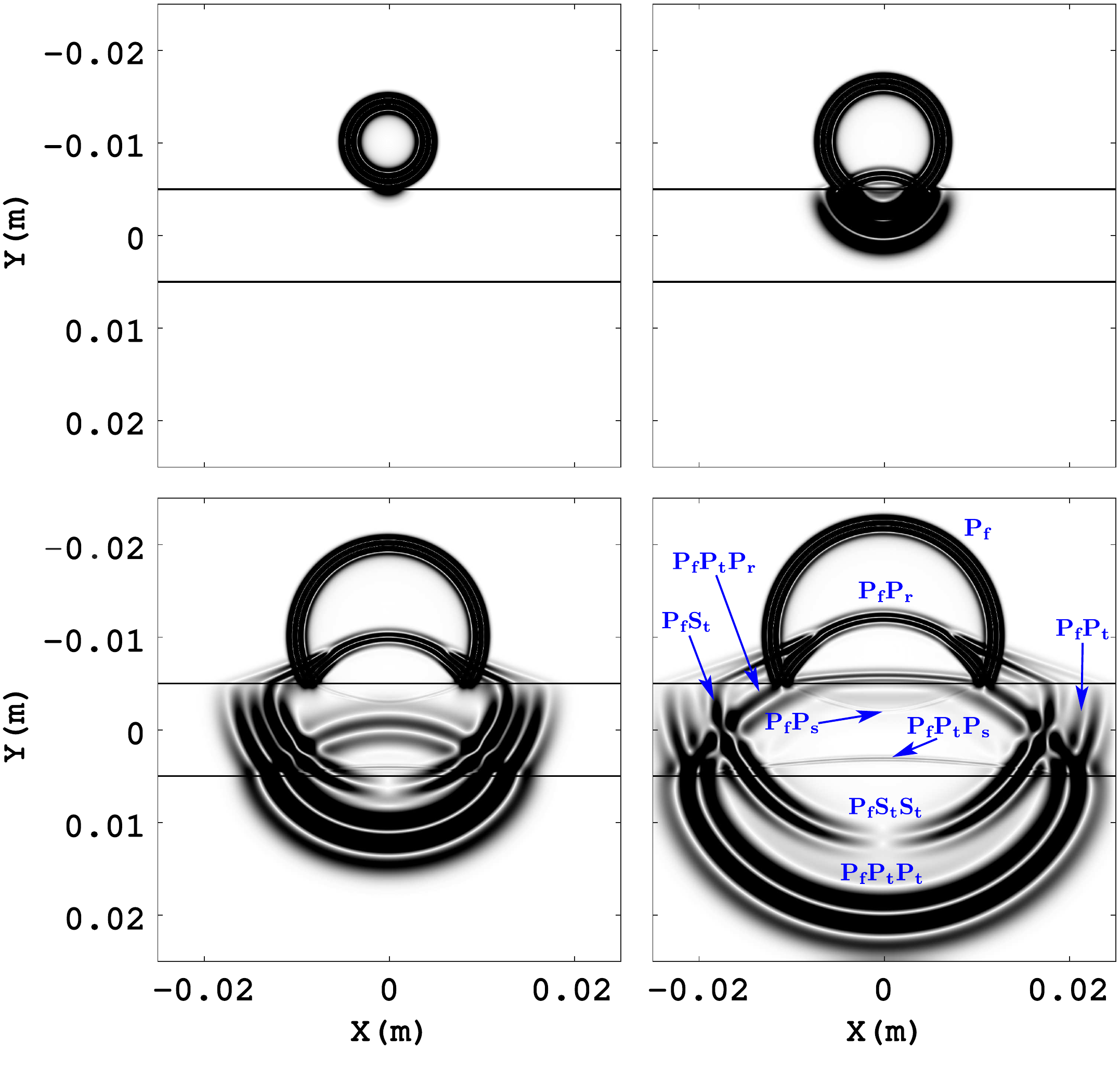}
	\end{center}
	\caption{{
	Snapshots of the norm of mixture velocity $ ||v||^2 $ for a layered medium: water (upper 
	layer, $ \phi=1 $), poroelastic (middle layer,  $ \phi=0.2 $) and elastic 
	solid  (bottom layer, $ \phi=0 $).}}
	\label{fig: layered_snap}
\end{figure*}

In order to present all types of waves, let us consider snapshots of the total 
velocity vector  for different moments of time. We strongly amplified the 
wavefield amplitude in Fig.\,\ref{fig: layered_snap} to be able to pick out the 
slow compressional waves in the snapshots. In order to interpret the waves arising 
in the medium, we use  'P' to mark P-waves  and  'S' to 
mark S-waves. The subscript 'r'  indicates the reflected waves, while 
the subscript 't'  indicates the transmitted waves. Also, we use the 
subscript 's' to identify a slow P-wave and the subscript 'f' to identify a 
pressure wave in the fluid layer.

The source in the water layer excites a pressure wave ( denoted by P$_\textrm{f}$ in 
Fig.\,\ref{fig: 
layered_snap}) which propagates towards the poroelastic layer.
This wave then reflects from the bottom of the water-poroelastic interface 
(P$_\textrm f$P$_\textrm r$) and generates a fast transmitted P-wave 
(P$_\textrm f$P$_\textrm t$), a slow 
transmitted P-wave (P$_\textrm f$P$_\textrm s$), and a transmitted 
S-wave (P$_\textrm f$S$_\textrm t$) in the  poroelastic medium. Afterwards, 
these waves generate a 
family of transmitted and reflected waves (including a transmitted P-wave 
(P$_\textrm f$P$_\textrm t$P$_\textrm t$), an S-wave 
(P$_\textrm f$S$_\textrm t$S$_\textrm t$), and a
reflected slow P-wave (P$_\textrm f$P$_\textrm t$P$_\textrm s$)) from the  
upper and lower boundaries of 
the poroelastic layer. 

In this example we see that the propagation of all types of waves 
predicted by the elasticity and poroelasticity theories is correct: slow waves arise only 
in the poroelastic layer, only pressure waves propagate in the liquid layer, while  
longitudinal and shear waves appear in the elastic medium.

\section{Conclusions}

An extension of the unified model of continuum fluid and solid mechanics 
\cite{DPRZ2016} for compressible fluid flows in elastoplastic porous media has 
been proposed.
The derivation  is based on the Symmetric Hyperbolic 
Thermodynamically Compatible (SHTC) theory \cite{SHTC-GENERIC-CMAT}, and the 
resulting model represents a combination of the unified continuum model from 
\cite{DPRZ2016} with the SHTC model for two-phase compressible flows from 
\cite{RomDrikToro2010}.
The governing equations satisfy two laws of thermodynamics (energy 
conservation and non-decreasing of  entropy) and form a first-order symmetric  
system 
which 
is hyperbolic in the sense of Friedrichs \cite{Friedrichs1958} if the 
generating thermodynamic potential is convex.

Based on the above-proposed non-linear model, a linearized first order PDE system 
for small-amplitude wave propagation in a stationary saturated porous medium 
has been derived. The linear system is written in terms of the velocity of the 
solid-fluid mixture, and the relative velocity of the phase motion, pressure and shear 
stress. Such a formulation allowed us a straightforward development of an 
efficient finite difference scheme on a staggered grid.  

A comparison of the above-proposed SHTC model and the classical Biot model for wave 
propagation in a saturated elastic porous  medium has been made at the aid of the 
dispersion 
analysis.
It turns out that, although the basic equations of the two models are different, 
the SHTC model is able to describe all the effects (in particular, the 
existence of a slow P-wave) predicted by Biot's theory, in good quantitative 
and qualitative agreement.

A number of two-dimensional test problems has been solved for the propagation of small-amplitude waves described by the formulated model. These test problems include, in particular, the study of the dependence of wavefields on model parameters.
The numerical results demonstrate that the SHTC model describes 
correctly all physical characteristics of the process.

Finally, we note that the developed poroelastic model is based on the SHTC formulation 
of 
mixtures. It thus should be kept in mind that other approaches to obtain continuum models for 
mixtures are possible \cite{BlokhinDorovsky1995,PKG-Book2018,Pavelka2014b}. In particular, on the 
PDE 
level, the 
continuum formulation for mixtures obtained in the GENERIC (General Equation for Non-Equilibrium 
Reversible-Irreversible Coupling) framework \cite{PKG-Book2018,Pavelka2016} differs from the SHTC 
formulation for mixtures used in this paper by a term $ \sim E_{w^l}\left(\frac{\partial 
w^k}{\partial 
x_l}-
\frac{\partial w^l}{\partial x_k}\right) $ missing in equation \eqref{eqn.relvelMS}. Also, on 
the physical level, the GENERIC formulation may differ by a different interpretation of the 
state variables, e.g. see a discussion on the SHTC compatible and alternative Poisson brackets for 
heat conduction  and the total momentum definition in \cite{SHTC-GENERIC-CMAT}. Nevertheless, 
for the 
case of small amplitude waves, such differences should not result in a different linear system and, 
in particular, the extra term mentioned above 
should vanish similar to the term $ v^l\left(\frac{\partial w^k}{\partial 
	x_l}-
\frac{\partial w^l}{\partial x_k}\right) $, see equation \eqref{stress.velocity.w}. Perhaps, 
differences in 
solutions should appear in case of finite deformations of poroelastic continuum. We hope to 
investigate this in detail in future publications.

\section*{Acknowledgements}
The authors are grateful to M. Yudin for valuable help in the manuscript 
preparation.
The research of E.R. and G.R. in Sects.2-4 was supported by the Russian Science Foundation  under grant 19-77-20004, the research in Sect.5 was supported by the Russian Foundation for Basic Research under grant 19-01-00347.
I.P. gratefully acknowledges the support of Agence Nationale de la Recherche (FR) 
(grant ANR-11-LABX-0040-CIMI) under program ANR-11-IDEX-0002-02.
The work of M.D. and I.P. was partially supported by the European Union's Horizon 2020 Research and 
Innovation  Programme under project \textit{ExaHyPE}. 
M.D. and I.P. also gratefully acknowledge funding from the Italian Ministry of Education, 
University and Research (MIUR) under the Departments of Excellence Initiative 2018--2022 attributed 
to DICAM of the University of Trento, as well as financial support from the University of Trento 
under the  \textit{Strategic Initiative Modeling and Simulation}. I.P. has further received funding 
from  
the University of Trento via the \textit{UniTN Starting Grant initiative}.

\noindent \section*{In memoriam}
\noindent This paper is dedicated to the memory of Dr. Douglas Nelson Woods ($^*$January 11\textsuperscript{th} 1985 - $\dagger$September 11\textsuperscript{th} 2019),
promising young scientist and post-doctoral research fellow at Los Alamos National Laboratory.
Our thoughts and wishes go to his wife Jessica, to his parents Susan and Tom, to his sister Rebecca and to his brother Chris, whom he left behind.

\printbibliography

\end{document}